\documentclass[12pt]{article}

\usepackage{graphicx}

\usepackage{scicite}
\usepackage{times}
\usepackage{amsmath}
\usepackage{amsfonts}
\usepackage{amssymb}

\newenvironment{sciabstract}{%
\begin{quote} \bf}
{\end{quote}}

\usepackage{float}

\makeatletter
\let\saved@includegraphics\includegraphics
\AtBeginDocument{\let\includegraphics\saved@includegraphics}
\renewenvironment*{figure}{\@float{figure}}{\end@float}
\makeatother

\newcommand{\beginsupplement}{%
        \setcounter{table}{0}
        \renewcommand{\thetable}{S\arabic{table}}%
        \setcounter{figure}{0}
        \renewcommand{\thefigure}{S\arabic{figure}}%
        \setcounter{section}{0}
        \renewcommand{\thesection}{S-\arabic{section}}
     }

\usepackage{mathtools}

\usepackage{dcolumn}
\usepackage{bm}
\usepackage{amsfonts}
\usepackage{booktabs}
\usepackage{siunitx}
\usepackage{amsmath,amssymb}
\usepackage{fdsymbol}
\usepackage{xcolor}
\usepackage{pifont}
\usepackage{bbding}
\usepackage{dirtytalk}

\usepackage{array}

\usepackage{setspace}
\usepackage{caption}
\title{A Novel Continuous Drop-Wise Condensation Technology for Improved Heat and Mass Transfer Efficiencies}

\author
{Ali Alshehri,${}^{1,2\ast}$ Jonathan P. Rothstein,${}^{3}$ H. Pirouz Kavehpour${}^{1}$\\
\\
\normalsize{${}^{1}$Mechanical and Aerospace Engineering Department,}\\
\normalsize{University of California, Los Angeles, CA 90095, USA,}\\
\normalsize{${}^{2}$Mechanical Engineering Department,}\\
\normalsize{King Fahd University of Petroleum and Minerals (KFUPM), Dhahran 31261, Saudi Arabia,}\\
\normalsize{${}^{3}$Department of Mechanical and Industrial Engineering,}\\
\normalsize{University of Massachusetts Amherst, Amherst, Massachusetts 01003-2210, USA.}\\
\\
\normalsize{$^\ast$To whom correspondence should be addressed; E-mail:  aalshehri@ucla.edu.}
}

\date{}

\begin{document}

\baselineskip24pt

\maketitle

 \clearpage
\section*{Abstract}
\begin{sciabstract}
Drop-wise condensation (DWC) has been the focus of scientific research in vapor condensation technologies since the 20th century. Improvement of condensation rate in DWC is limited by the maximum droplet a condensation surface could sustain. Furthermore, The presence of non-condensable gases (NCG) reduces the condensation rate significantly. Here, we present continuous drop-wise condensation (CDC) to overcome the need of hydrophobic surfaces while yet maintaining micron-sized droplets. By shifting focus from surface treatment to the force required to sweep off a droplet, we were able to utilize stagnation pressure of jet impingement to tune the shed droplet size. The results show that droplet size being shed can be tuned effectively by tuning the jet parameters. our experimental observations showed that the effect of NCG is greatly alleviated by utilizing our technique. An improvement by at least six folds in mass transfer compactness factor compared to state-of-the-art dehumidification technology was possible. 
\end{sciabstract}
 \clearpage

\section*{Introduction}

Condensation is a perplexing problem to fully uncover, yet, its applications play crucial roles in industrial development \cite{jadhav2012,enright2014dropwise,wilson2014principles,zhang2015breath,warsinger2015entropy}. In order to promote more efficient applications, improving the condensation process has been the focus of many scientific research. Various condensation heat transfer augmentation techniques have been utilized, these techniques can be classified to; Passive \cite{miljkovic2016,chen2017,Li2018,zhao2018,famileh2017,datt2018,rashidi2018}; active \cite{Shahriari2017,Miljkovic2013DWCjump}; and compound techniques. Unlike passive, active techniques require external forces to be applied either on the heat transfer surface or working fluid. Unsurprisingly, passive techniques have caught major attention of the scientific community because of their easier implementation in existing industrial applications. Contrarily, active techniques require additional equipment to exert the required forces to improve the heat and mass transfer efficiencies. This, in turn, means higher capital expenditures and operating expenses. Observing the state-of-the-art literature, the most trending technology is altering surface wettability \cite{miljkovic2016}. However, it is still under development due to the chemical and physical durability issues of coating techniques \cite{daniel2013,miljkovic2016,cho2017,ahlers2019dropwise,ma2020recent}.

Major research efforts have been focused on developing durable and cost effective coating techniques to promote drop-wise condensation (DWC) and mitigate the formation condensate films on surfaces, i.e. film-wise condensation (FWC). DWC has shown to possess at least an order of magnitude improved heat transfer coefficient compared to FWC \cite{rose2002dropwise}. This improvement is highly dependant on the frequency of droplet shedding on surfaces. Droplets shedding has been achieved primarily by gravity assistance \cite{rose2015,kim2011dropwise,dimitrakopoulos1999gravitational}, droplet jumping \cite{miljkovic2012jumping,Miljkovic2013DWCjump,yuan2019enhanced,peng2020breaking}, drag force \cite{hu2013displacement,roisman2015dislodging,seevaratnam2010laminar,milne2009drop,razzaghi2018shedding}, or by capillary driven movement \cite{yan2020near,li2018spontaneous}. It has been widely accepted that droplets of diameters below 20 micron contribute about 80\% of the total heat transfer to the surface \cite{graham1973drop}. This implies that removing droplets of higher diameters is preferred, otherwise, their higher thermal resistance and coverage area will impede further condensation. Combining superhydrophobic surfaces and a shedding mechanism might seem to be the obvious solution for achieving efficient DWC. However, superhydrophobic surfaces are characterized by their low nucleation site density for condensation and higher droplet thermal resistance, therefore presenting a conflicting purpose \cite{ma2020recent}. Therefore, there is a high demand to sustain efficient DWC with innovative techniques that go beyond surface modifications.

An extremely important concern most passive augmentation techniques has not resolved is the degradation of heat/mass transfer coefficients caused by the existence of non-condensable gases (NCG) \cite{ma2020recent,alshehri2020numerical}. Experimentally, degassing prior to running condensers has been successful in alleviating the effect of NCG \cite{rose2015,ma2020recent}. Despite the experimental convenience of such method, it is a highly impractical solution in large scale condensers. NCG can find their way into condensers via leak points or as chemical reaction products of vapor interacting with the equipment material \cite{Zhang2017}. On another front, the emerging humidification-dehumidification (HDH) desalination technology relies heavily on NCG as carrier gases. The premise of this technology is the low energy required to humidify air compared to other thermal desalination counterparts \cite{Narayan2010,Giwa2016,srithar2018}. Even though it is evident that the dehumidifier in HDH technology is highly inefficient, the heat transfer deficit has been compensated by three alternatives. They are; (1) extended contact area \cite{chafik2004,chafik2003,farid2003,chang2014experimental} (2) direct contact between humid air and cooling medium \cite{klausner2006,dawoud2006,hu2011,agboola2015,tow2014,liu2016,sadeghpour2019water}; and (3) different NCG carrier \cite{narayan2011,arabi2003}. Even though the former two solutions are promising, the latter seems to address the problem at its core, i.e. the effect of vapor diffusion coefficient thus condensation rate. Therefore, there is a pressing demand on working out a solution to enhance condensation with the presence of NCG.

To overcome the problem of sustaining efficient condensation without requiring unstable and expensive surface modifications, an active method needs to be designed. The active augmentation method needs to mitigate the negative effect of NCG while maintaining practical applicability. Several active methods have been tried, such as fluid/surface vibration \cite{Raben1961,chen2017droplet,migliaccio2014resonance,moradi2020vibration}, electrohydrodynamic effects \cite{velkoff1965,Baratian2018,Dey2018,Miljkovic2013DWCjump,Traipattanakul2019}, and rotating surfaces \cite{mohamed2006effect,peng1998theoretical,yanniotis1996experimental}, to name a few. The general goal of the different active methods is to prevent the condensate from growing by actively sweeping it off the surface. While this has shown to be effective, surface wettability is still important to generate DWC rather than FWC. In addition, the effect of NCG is still not resolved with the aforementioned methods. Here, we investigate utilizing jet impingement as an active method for providing DWC on surfaces with varying surface wettability. The jet impingement method not only helps with shedding droplets on wettable surfaces but also helps with mitigating the effect of NCG. The utilization of jet impingement in heat and mass transfer applications has been studied in heating/cooling for single phase flow \cite{viskanta1993,lienhard2006}, drying application \cite{mujumdar201415}, nucleate boiling \cite{qiu2015recent}, and spray cooling \cite{khangembam2020experimental}. Recently, on-demand impingement of pure steam jet has shown to alleviate the effect of NCG in accidental leakage \cite{ji2019enhancement,ji2020effective}. We recently showed that utilizing jet impingement in condensing water vapor in a humid environment results in a breath figure spot\cite{ALSHEHRI2021121166}. This spot defines the boundary over which effective condensation takes place.

To sustain efficient DWC without requiring unstable and expensive surface modifications, we present a novel `continuous drop-wise condensation' (CDC) as a method to tune the maximum droplet size on modified and unmodified condensation surfaces. We also present CDC as a method to improve condensation with NCG by means of thinning the diffusion boundary layer and therefore reducing the resistance to diffusion. Impinging a modified or unmodified surface with a jet of humid air or pure vapor results not only in a higher heat and mass transfer coefficients but provides an excellent droplet shedding mechanism (Fig.1). Controlling the diameter of droplet shedding is made possible by tuning the jet parameters, e.g. exiting diameter, velocity, and standoff distance. To provide evidence of the proposed mechanism, several experiments were conducted under various jet parameters as well as different surfaces with a wide range of advancing contact angles, i.e. $\theta_A$ = 70$^o$ - 160$^o$. In addition, we utilize our experimental observation to show that CDC provides over 6-fold improvement in compactness factor compared to state-of-the-art dehumidifiers. Furthermore, using an analytical model, we show that CDC provides enhancement in heat flux of over 300\% compared to gravity-assisted shedding mechanisms. This is made possible by the improved mechanism of tuning the maximum droplet size compared to state-of-the-art techniques. Finally, We provide a theoretical framework for understanding droplet dynamics by comparing the different forces acting on a droplet during jet impingement.

\begin{figure}[h]
\centering
\includegraphics[width=\textwidth]{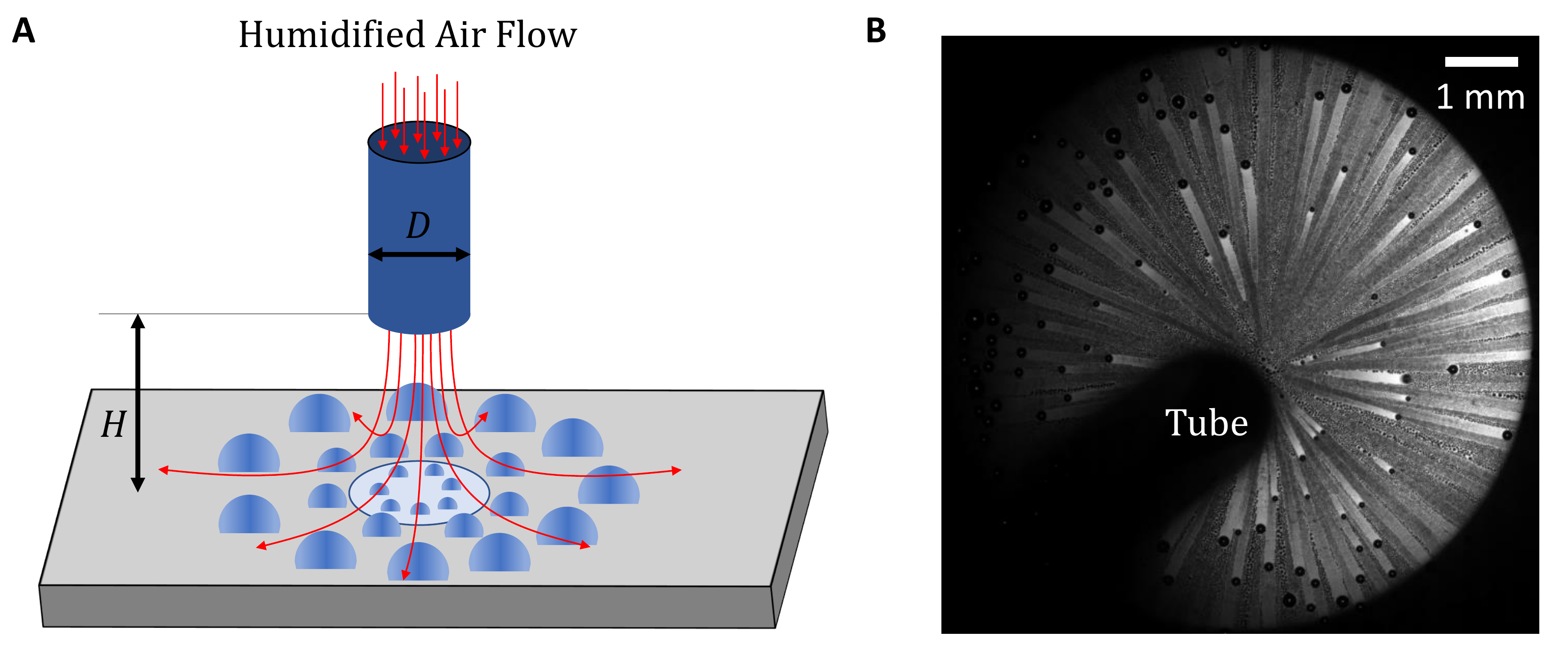}
\caption{ \textbf{ Utilizing Jet impingement as a means of continuous drop-wise condensation (CDC).} \textbf{A.} A schematic of CDC illustrating the condensation mechanism utilizing jet impingement as a means of enhanced condensation rate and droplet shedding. pure vapor or Humidified-air jet exits a tube of diameter ($D$) at a standoff distance ($H$) with a mean velocity ($v_j$). \textbf{B.} A still Microscopic image of CDC under a selected experimental condition ($D$ = 0.047 inches, $H$ = 0.32 inches and $Re_j$ = 3600). Complete description of experimental set up is presented in methods section and supplementary material Figure S1}
\label{Fig1}
\end{figure}

\section*{Results}

\subsection*{Shedding of droplets under jet impingement}

The dynamics of jet impingement on a surface is characterized by a stagnation region that spans almost two nozzle diameters \cite{jambunathan1992,hollworth1983entrainment}. Beyond this radial location, a wall jet forms that behaves similar to a Blasius boundary layer. Here we show that the force within the stagnation region provides an excellent shedding capability. To study this, we visualize under a microscope the growth and onset of shedding of droplets under different jet Reynolds numbers $Re_j = 4 Q / \pi \nu D$, where $Q$ is the jet flow rate, $\nu$ is the kinematic viscosity of humid air, and $D$ is the tube exit diameter (Fig.2). To provide consistent comparisons, the surfaces were cleaned prior to each experimental run as outlined in the methods section. Additionally, the advancing and receding contact angles were measured before and after each experimental run with no significant changes due to mobile droplet shearing effects. Here, we utilized a hydrophobic Si wafer ($\theta_A$ = 107$^o$ and $\theta_R$ = 103$^o$) as the condensation surface (see methods section and Table 1). The jet flow rate was first set to the desired value of jet mean velocity, after which the surface temperature was brought down to the desired temperature ($T_s$ = 15$^oC$). A high speed camera (Photron, FASTCAM Nova) attached to an optical microscope (Nikon, AZ100) was utilized to obtain videos and images of the condensation process (Video 3 and Fig.2). 

\begin{figure}
\centering
\includegraphics[width=0.62\textwidth]{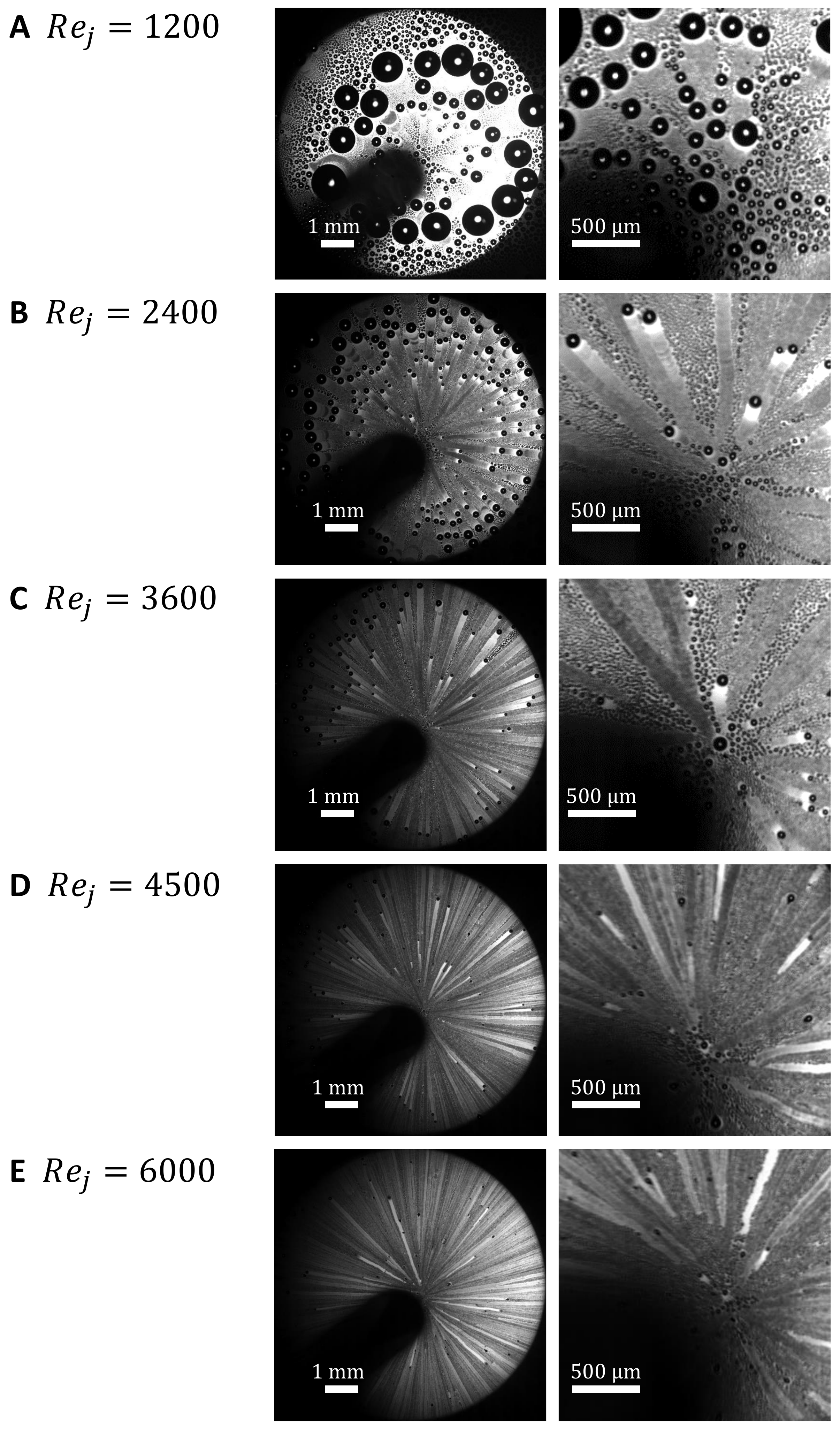}
\caption{ \textbf{Effect of jet Reynolds number on the size of shed droplets.} Images show results of condensation experiments performed at an ambient temperature of 21$^o$C and surface temperature of 15$^o$C where the relative humidity of ambient and jet were 60\% and 95\%, respectively. The tested surface was a hydrophobic Si wafer with $\theta_A$ = 107$^o$ and $\theta_R$ = 103$^o$. Two different close up view are shown for condensation with jet Reynolds numbers of \textbf{A.} $Re_j=$ 1200, \textbf{B.} $Re_j=$ 2400, \textbf{C.} $Re_j=$ 3600, \textbf{D.} $Re_j=$ 4500, and \textbf{E.} $Re_j=$ 6000.  }
\label{Fig2}
\end{figure}

In Fig.2, we show the quasi-steady droplet distribution on the condensation surface under different jet Reynolds numbers. In these experiments, the relative humidity of ambient air was 60\% and the jet was 95\%. The tested surface was a hydrophobic Si wafer with an advancing contact angle of $\theta_A$ = 107$^o$ and a receding contact angle of $\theta_R$ = 103$^o$. In Fig.2 and the corresponding Video 3 found in the supplemental material, the droplets can be observed to go through three main growth stages with time: Stage I - slow growth of stationary drops beneath the impinging jet; Stage II - fast growth as mobile droplets move radially outward merging with other droplets; and Stage III - slow growth of droplets that have come to rest far from the center of the impinging jet. In Stage I, single droplets residing on nucleation sites present on the surface initially start growing by direct condensation on their exposed surfaces. As single droplets reach a critical size ($R_c \sim 1/\sqrt{4 N_s}$, where $N_s$ is the nucleation site density), they start coalescing with neighboring droplets. The growth mechanism in Stage I can be described by a combination of direct condensation and coalescence with neighboring droplets. As droplets grow to another critical size, determined by the jet shedding capability, they start departing their equilibrium location on the surface. The onset of droplet departure is observed to be located near the stagnation region. This implies that droplets in this region possess higher growth rates and experience higher drag forces compared to droplet located further away. It is also observed that the critical droplet size at the onset of departure is reduced with increasing the jet Reynolds number. As droplets depart their first equilibrium location, they start moving radially outward coalescing with droplets in their path (Stage II). The action of movement and coalescence acts as an effective sweeping mechanism for improved DWC. The growth mechanism of a mobile droplet is determined only by coalescence and negligible direct condensation. After droplets move to locations further from the center of the impinging jet, the retention force due to surface tension overcomes the drag force by the gas flow around them and they decelerate and stop. Stationary far-field droplets can be observed clearly in Fig.2A but are out of frame in Fig.2B-C. Stationary droplets continue to grow by smaller mobile droplets that are generated from the stagnation region following their mother droplet path (stage III). 

It is also worth noting that as droplets move along their radial path, they leave dry traces which appear as white traces under the microscope. The process of droplet growth then repeats on the traces until another droplet coming from the stagnation region sweeps it away. Under ideal circumstances, the size of the stationary droplets withing these traces is limited by the size of the mobilized droplets. Hence, a mobilized droplet can be thought of as the maximum size a droplet can grow. In some circumstances, surface imperfections or dust particles can interfere with the mobile droplet sweeping action but could be mitigated by cleaning the surface thoroughly.

In Fig.2A-E, a magnified view of the droplets formed beneath the impinging jets are shown to illustrate the effect that jet velocity and jet Reynolds number have on droplet mobility.  At the smallest Reynolds number presented in Figure 2A, $Re_j = 1200$, droplets were not observed to shed even as they grew quite large.  In Fig.2B, at a $Re_j = 2400$, droplets with radii greater than $R_{max}$ = 33 $\mu$m were observed to shed and move radially outward coalescing with smaller drops and growing as they moved. With increasing Reynolds number, a further reduction in shedding drop size was observed.  For the highest flow rate tested, $Re_j = 6000$, droplets with radii greater than $R_{max}$ = 13 $\mu$m were observed to shed. These drop sizes are significantly smaller than the case of gravity-assisted droplet shedding or the shearing effect of boundary layer flows where only droplets with radii above 250 microns shed from a hydrophobic surface \cite{yan2020near}. Our results clearly demonstrate that the maximum condensed droplet size can be efficiently tuned by controlling the impinging jet velocity and Reynolds number. 

Surface wettability is an important factor in determining the shedding capability of a surface regardless of the active mechanism generating the shedding forces. In order to test the effect of wettability on CDC, the results of a series of experiments are presented in Fig.3 for five different surfaces with a wide range contact angles at a fixed jet Reynolds number of $Re_j$ = 3600.  In these experiments, ambient temperature was 21$^o$C and the surface temperature 15$^o$C while the relative humidity of ambient air and the jet were 60\% and 95\%, respectively. The advancing, $\theta_A$, and receding contact angle, $\theta_R$, for each surface are presented in Table 1 along with the contact angle hysteresis, $\theta_A - \theta_R$. The condensation process and shedding capability is visualized in Fig.3 with videos available as supplementary material (Videos 1-5). Two hydrophilic surfaces, one smooth and one microstructured, with different contact angles are presented in Fig.3A and 3B. On the smooth hydrophilic surface, Fig.3A, droplets with radii greater than 20 $\mu$m were observed to shed. The microstructured hydrophilic surfaces has roughly twice the contact angle hysteresis of the smooth hydrophilic surface, $\theta_A - \theta_R = 20^o$ vs $13^o$.  As a result, the mobility of the droplets is hindered by the increased interfacial pinning force on the droplets caused by the presence of the microstructures and an increase in the radius of the shedding drops was observed to a value of 36 $\mu$m. On the other hand, minimizing the contact angle hysteresis, as is done for both the hydrophobic and the nanostructured superhydrophobic surface shown in Fig.3C and 3E, dramatically reduces the minimum droplet shedding radius by reducing the interfacial pinning force.  For example, droplet radius of the drops shedding from the hydrophobic surface in Fig.3C was 13 $\mu$m.  For the nanostructured superhydrophobic surface shown in Fig.3E, the surface looks clear under the microscope with no evidence of the pathlines of shedding droplets clearly visible in Fig.3A-C. This is probably due to the low condensation rate due in large part to the low density of nucleation sites on these nanostructured superhydrophobic surfaces, but it could also be the result of droplet jumping from the surface as they coalesce and interfacial energy is recovered in the form of kinetic energy.  Some evidence for the presence of droplet jumping can be seen in the videos provided in the supplementary materials.  Droplet jumping has been shown to improve the heat transfer to a surface during condensation \cite{miljkovic2012jumping} and will be discussed in more detail later.  Finally, we analyze the results of the microstructured superhydrophobic surface in Fig.3D. Interestingly, even though it had the largest advancing contact angle, the microstructured superhydrophobic surface also had the highest contact angle hysteresis.  The CDC experiments for this were characterized by a significant pinning of droplets and a very large variability and uncertainty in the size of the shedding droplets. As a result, the discussion of the drop dynamics that follows will be focused primarily on Surfaces 1, 2 and 3 for which repeatable data could be obtained.

\begin{table}
\caption{Advancing and Receding contact angles of the different surfaces used. Surfaces have different wettability and contact angle hysteresis. \\}
\centering
\renewcommand{\arraystretch}{0.6}
\begin{tabular}{c c c c c}
\toprule
\normalfont{Name: Description} & $\theta_A$ & $\theta_R$ & $\theta_A - \theta_R$ &  $cos(\theta_A) - cos(\theta_R$)     \\
\midrule
\normalfont{Surface 1: Hydrophilic} & $85\pm 2^o$ & $72\pm 2^o$ & $13\pm 3^o$ & $0.22\pm 0.05$  \\
\normalfont{Surface 2: Hydrophilic Microstructured} & $70\pm 2^o$ & $50\pm 2^o$ & $20\pm 3^o$ & $0.30\pm 0.04$  \\
\normalfont{Surface 3: Hydrophobic} & $107\pm 2^o$ & $103\pm 2^o$ & $4\pm 3^o$ & $0.03\pm 0.05$  \\
\normalfont{Surface 4: Superhydrophobic Microstructured} & $160\pm 2^o$ & $127\pm 2^o$ & $33\pm 3^o$ & $0.34\pm 0.03$  \\
\normalfont{Surface 5: Superhydrophobic Nanostructured} & $157\pm 2^o$ & $154\pm 2^o$ & $3\pm 3^o$ & $0.022\pm 0.02$  \\
\bottomrule\\
\end{tabular}
\label{table:contact_angles}
\end{table}

\begin{figure}
\centering
\includegraphics[width=0.65\textwidth]{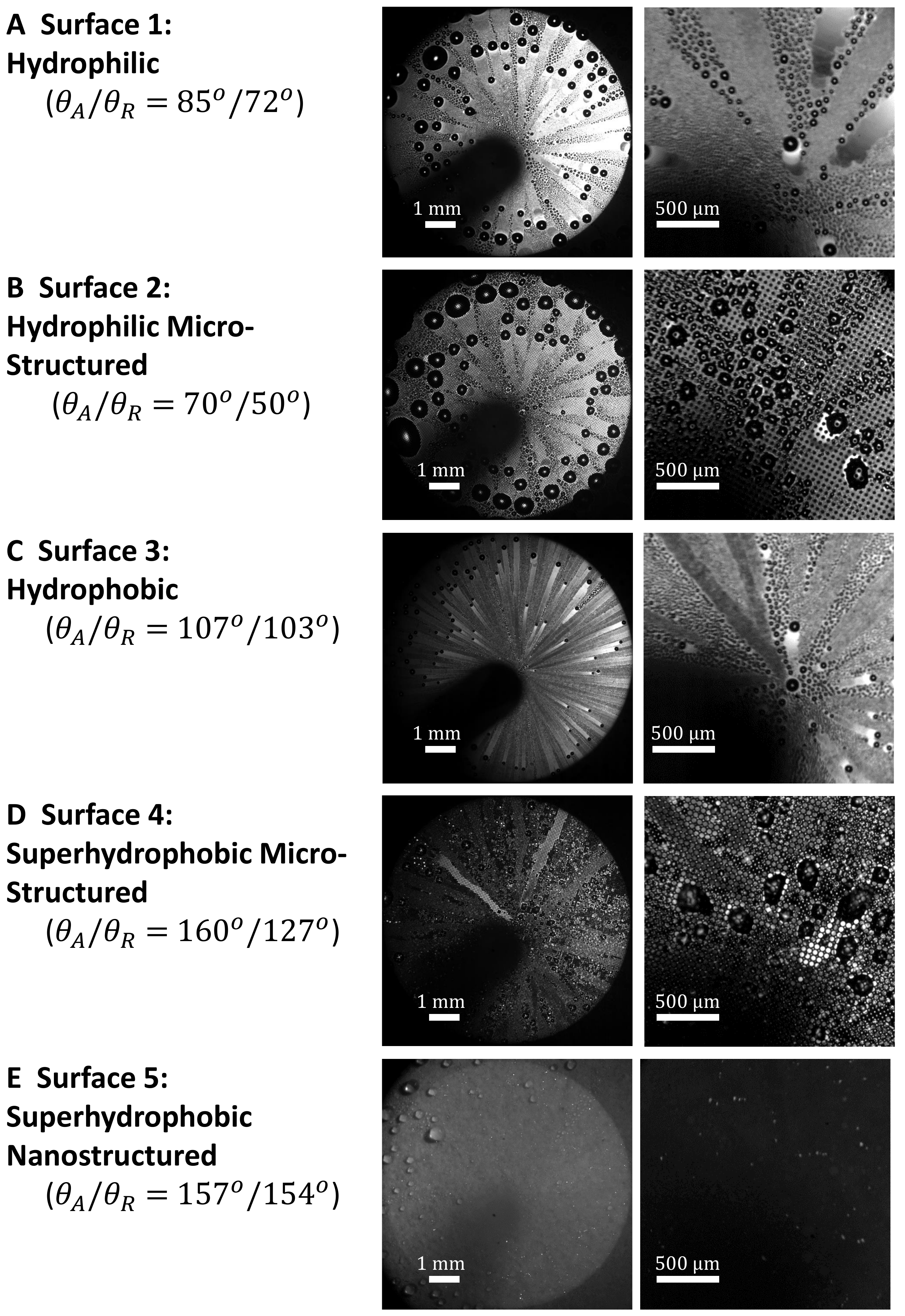}
\caption{ \textbf{Effect of surface wettability on the condensation process.} The condensation experiments were performed with an ambient air temperature of 21$^o$C and surface temperature of 15$^o$C, while the relative humidity of ambient and jet were 60\% and 95\%, respectively. Two different close up views are shown for condensation with varying surface wettability and contact angle hysteresis at a fixed jet Reynolds number of $Re_j$ = 3600. The advancing and receding contact angles is displayed beside images of each surface.}
\label{Fig3}
\end{figure}


\clearpage
\subsection*{Dehumidification with CDC for improved heat and mass transfer rates}


In this section, we analyze the heat transfer and condensation rates for the CDC and compare their performance with the current state of the art dehumidifiers.  To compare the heat flux or condensation rate across many different dehumidifier designs, we compare their compactness factors. The compactness factor indicates the heat or mass transfer rate per unit driving potential (temperature or vapor mass fraction) per unit volume and is defined as follows \cite{sadeghpour2019water}.
\begin{equation}
    C_h = \frac{h A}{V}
\end{equation}
\begin{equation}
    C_m = \frac{h_m A}{V}
\end{equation}
where $C_h$ is the compactness factor of heat transfer unit, $C_m$ is the compactness factor of mass transfer unit, $h$ is the heat transfer coefficient, $h_m$ is the mass transfer coefficient, $A$ is the surface area over which measurement takes place, and $V$ is the volume of the dehumidification system. In order to experimentally evaluate the mass transfer coefficient, we utilized an optical method of observing the growth of condensate droplets near the impingement region. The mass transfer coefficient can be written as follows.
\begin{equation} \label{local}
    h_m =  \frac{\rho_l}{(\omega_\infty - \omega_s) A} \frac{dV_d}{dt} 
\end{equation}
where $\rho_l$ is the liquid density, $\omega_\infty$ is the vapor mass fraction at ambient conditions, $\omega_s$ is the vapor mass fraction evaluated at the surface temperature and $dV_d/dt$ is the condensate volumetric growth per unit time. The volumetric growth rate can be calculated directly from the video images.  To calculate the droplet volume, the droplets are assumed to take the form of a spherical cap because their radii are smaller than the capillary length. For a droplet that is a spherical cap with an optically observed radius of $R$ on the hydrophobic surface, the volume of a droplet can be calculated from
\begin{equation}
    V_d = 
    \frac{\pi}{3} (2 + cos \theta_A)(1 - cos \theta_A)^2 R^3
\end{equation}
The evolution of droplet volume with time was calculated within the impingement region (a surface with a diameter of 1 mm).  Additionally, the number and volume of droplets shedding and leaving the impingement region was tracked with time.  Because droplets departing the impingement region collect more liquid as the travel radially outward, this procedure provides the lower limit of the condensation rates and the mass transfer coefficient (Fig.4A).

\begin{figure}[h]
\centering
\includegraphics[width=\textwidth]{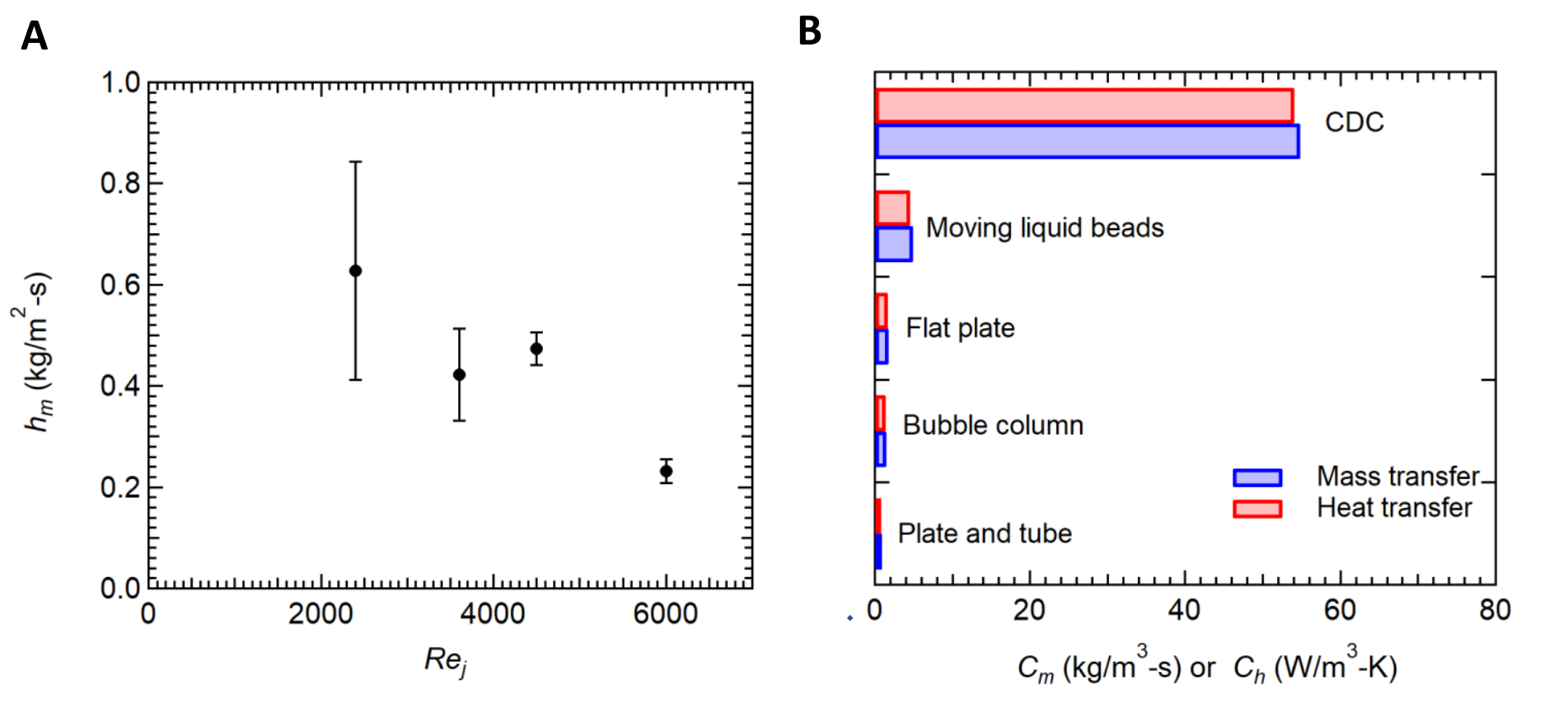}
\caption{ \textbf{Heat and mass transfer rate improvement of CDC.} \textbf{A.} Experimental evaluation of mass transfer coefficient, $h_m$ of CDC as a function of impinging jet Reynolds number. The experiments were performed at an ambient temperature of 21$^o$C and surface temperature of 15$^o$C where the relative humidity of ambient and jet were 60\% and 95\%, respectively. The tested surface was a hydrophobic Si wafer with $\theta_A$ = 107$^o$ and $\theta_R$ = 103$^o$. \textbf{B.} Comparison of heat and mass transfer compactness factors for different state-of-the-art dehumidifiers; plate-and-tube dehumidifier\cite{farid2003}, bubble column dehumidifier\cite{tow2014}, flat plate dehumidifier \cite{sievers2013design}, moving liquid beads dehumidifier \cite{sadeghpour2019water}. The average value of our current experiments is shown.
}
\label{Fig4}
\end{figure}

In Fig.4, results for the mass transfer coefficient and the mass transfer compactness factor are presented for different jet Reynolds numbers. In these experiments, the relative humidity of ambient air was 60\% and the jet was 95\%. The tested surface was a hydrophobic Si wafer with an advancing contact angle of $\theta_A$ = 107$^o$ and a receding contact angle of $\theta_R$ = 103$^o$. The mass transfer coefficient ranged from $0.2 < h_m < 0.6$ kg/m$^2$-s while the compactness factors obtained were in the range of $30 < C_m < 300$ kg/m$^3$-s. The largest values of compactness factor were found for the lowest Reynolds numbers tested.  In Fig.4B, a comparison of the compactness factor for different dehumidifiers along with the current CDC method is presented for condensation on hydrophobic surfaces. Four state-of-the-art dehumidifiers are shown namely (1) plate-and-tube dehumidifier \cite{farid2003}; (2) bubble column dehumidifier\cite{tow2014}, (3) flat plate dehumidifier \cite{sievers2013design}, (4) moving liquid beads dehumidifier \cite{sadeghpour2019water}. As seen in Fig.4B, even when compared against the lowest value obtained utilizing CDC, the compactness factor of CDC exceeds the highest state-of-the-art dehumidifier by almost 6 times. Thus, CDC provides an extremely compact dehumidifier as well as a very compact heat transfer technique. We expect that even higher values of compactness factor are possible through optimization of geometric parameters like nozzle standoff distance and impinging jet diameter.


\clearpage
\subsection*{Pure steam condensers with CDC for improved heat and mass transfer rates}


We now turn to the case of utilizing our method to improve the heat and condensation rates for the case of pure vapor/steam. To characterize the condensation process due to the CDC improved droplet shedding, we visualized the condensation process on different jet Reynolds numbers (Fig.2) and different surface wettability (Fig.3). As observed, when droplets grow to their maximum droplet size near the stagnation region, they are shed by the jet impingement action. In contrast, in regular gravity-assisted DWC, the shedding mechanism is mainly due to the weight of the droplet. This requires droplets to grow by direct condensation and coalescence with neighboring droplets until reaching the capillary length (mm range) beyond which their weight overcomes the surface tension force. For pure vapor condensation, the high thermal resistance of large droplets reduces the heat transfer significantly (Fig.S3). In addition, allowing a surface to sustain high droplet size before shedding results in a decrease in the small droplet density (Fig.5A). It is well established that droplets with radii below 20 $\mu m$ contribute to almost 80\% of the overall heat transfer to the surface \cite{graham1973drop}. Therefore, CDC is an efficient means of reducing the maximum droplet size and increasing the population density of droplets below 20 $\mu m$ (Fig.5A).  

To evaluate the heat transfer flux, we utilize the analytical model developed by Rose et al. \cite{leFevre1966theory,kim2011dropwise}. The overall heat flux to a surface exposed to condensation in an ambience of pure vapor is given as
\begin{equation}
    q'' = \int_{r_{min}}^{r_e} q_d(r,\theta) n(r,\theta) dr + \int_{r_e}^{r_{max}} q_d(r,\theta) N(r,\theta) dr 
\end{equation}
where $q_d(r,\theta)$ is the heat transfer through a single droplet with Radius $r$ and contact angle $\theta$, $n(r,\theta)$ and $N(r,\theta)$ are the population density of of small and large droplets, respectively. Droplet below the critical droplet radius ($r_e = 1\sqrt{4N_s}$) grow by direct condensation (small droplets) while droplet above the critical radius grow by direct condensation an coalescence with neighboring droplets (large droplets). The heat transfer through a single droplet can be written as
\begin{equation} \label{q_ddd}
    q_d (r,\theta) = {\pi r^2 (T_{sat} - T_s - \frac{2 T_{sat} \gamma}{\rho_l h_{fg} r })}(\frac{1}{2 h_i (1 - \cos{\theta})} + \frac{r \theta}{4 k_l \sin{\theta}} + \frac{\delta_s}{k_s \sin^2{\theta}} )^{-1}
\end{equation}
Detailed analysis is given in Supplementary material (section S-4). It is worth noting that the previous equation was derived for the case of pure vapor. The main variable CDC introduces is the modification of maximum droplet radius value which consequently changes the heat transfer characteristics as shown in Fig.5B. Because for this case the jet advective transport does not introduce a thermal resistance, it acts only as an improved shedding mechanism and no further modification is required to Eq.\ref{q_ddd}. In Fig.5B, the heat flux to a surface is improved significantly by lowering the maximum droplet radius. Improvement as high as almost 150\% in heat flux (or condensation rate) can be obtained by utilizing a hydrophilic surface with maximum droplet radius of 20 $\mu m$. If one compares utilizing the hydrophilic surface over the superhydrophobic surface, a maximum improvement of heat flux above 375\% can be achieved (see supplementary material section S.4). This shows that heat transfer and consequently condensation rate can be improved by tuning the maximum droplet size which can be achieved easily with CDC.  

\clearpage

\begin{figure}
\centering
\includegraphics[width=\textwidth]{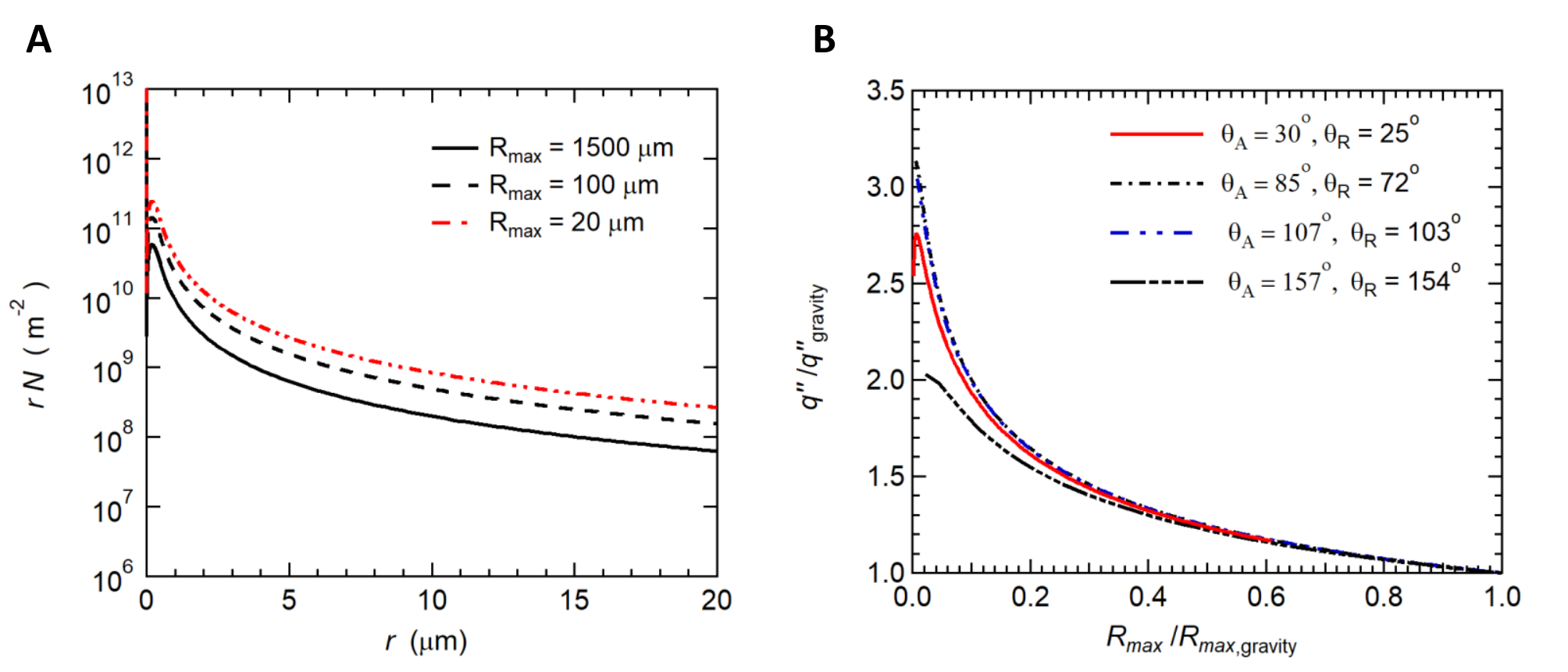}
\caption{\textbf{Heat transfer characterization of CDC.} \textbf{A.} The population density of droplets ($rN$) of radii below 20$\mu$m for different maximum droplet radius ($R_{max}$) being shed by the jet impingement action. The ordinate is defined as the number of droplets per unit surface area. \textbf{B.} Heat flux ($q''$) to a surface exposed to \underline{pure vapor} analytically evaluated at different maximum droplet radius ($R_{max}$). The heat flux and maximum droplet radius are normalized with values evaluated in case of gravity-assisted droplet shedding (see supplementary material section S-4).
}
\label{Fig5}
\end{figure}


\clearpage
\subsection*{Jet-droplet dynamics}

\begin{figure}
\centering
\includegraphics[width=\textwidth]{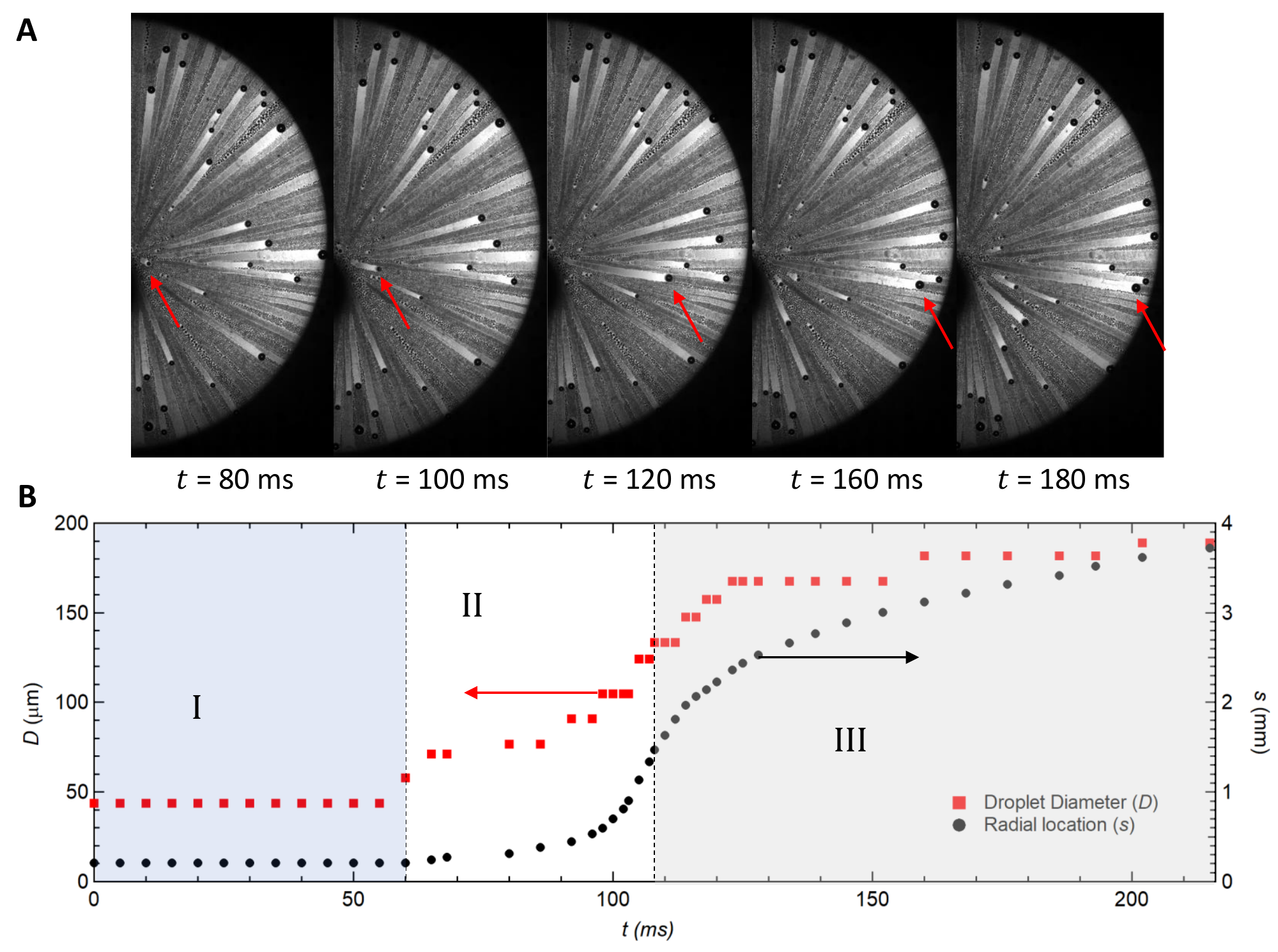}
\caption{\textbf{A typical time evolution of droplet motion and growth.} \textbf{A.} Time evolution of a droplet departing its first equilibrium location and moving radially outward. \textbf{B.} Typical transience plot illustrating both the location of the droplet ($s$) and its diameter ($D$) as it coalesces with other droplets in its path. The droplet goes through three different periods; (I) a waiting period before droplet departure, (II) an accelerating droplet period, and (III) a decelerating droplet period. This plot is generated for a selected case of $Re_j=3600$ and hydrophobic surface ($\theta_A$ = 107$^o$ and $\theta_R$ = 103$^o$). }
\label{Fig6}
\end{figure}

The dynamics of droplet shedding and motion is governed by an interplay between droplet inertia, droplet retention forces due to surface tension, viscous friction within the drop, and drag force due to flow of air around the drop. The effect of each of these was investigated by tracking the motion and size of individual droplets at the different stages of growth and motion as they as they moved across different substrates under different impinging jet conditions. In Fig.6A, a sample droplet is traced with time along the hydrophobic surface ($\theta_A$ = 107$^o$ and $\theta_R$ = 103$^o$) for one representative case at a jet Reynolds number of $Re_j=3600$. The diameter and location of the droplet is plotted as functions of time in Fig.6B. Three different periods can be clearly distinguished from the data: Period I - a waiting period before droplet departure; Period II - a period of radial acceleration of the droplet; and Period III - a period of droplet deceleration until the droplet comes to rest. The onset of droplet motion occurs after a waiting period during which the droplet grows through condensation to a critical size.  At this critical size, the aerodynamic drag force acting on the droplet becomes larger than the interfacial retention forces acting along the contact line between the droplet and the surface. After the droplet departs its initial location (onset of period II), it coalesces with droplets in its path resulting in a fast growth in droplet size and an  acceleration across the surface because with increasing size the aerodynamic drag force grows faster with droplet radius than then the interfacial retention force ($R^2$ vs. $R$).  During period II, the droplet roughly triples meaning the volume has increased by a factor close to thirty. During period III, the droplet decelerates and the rate of diameter growth slows as fewer coalescence event occur.  This deceleration occurs because the strength of the shear flow near the wall decreases as the drop moves radially outward from the center of the impinging jet ($1/s$).  Qualitatively similar results were observed for all surfaces tested provided the jet Reynolds number was larger than the critical Reynolds number to initiate droplet motion.

\begin{figure}
\centering
\includegraphics[width=\textwidth]{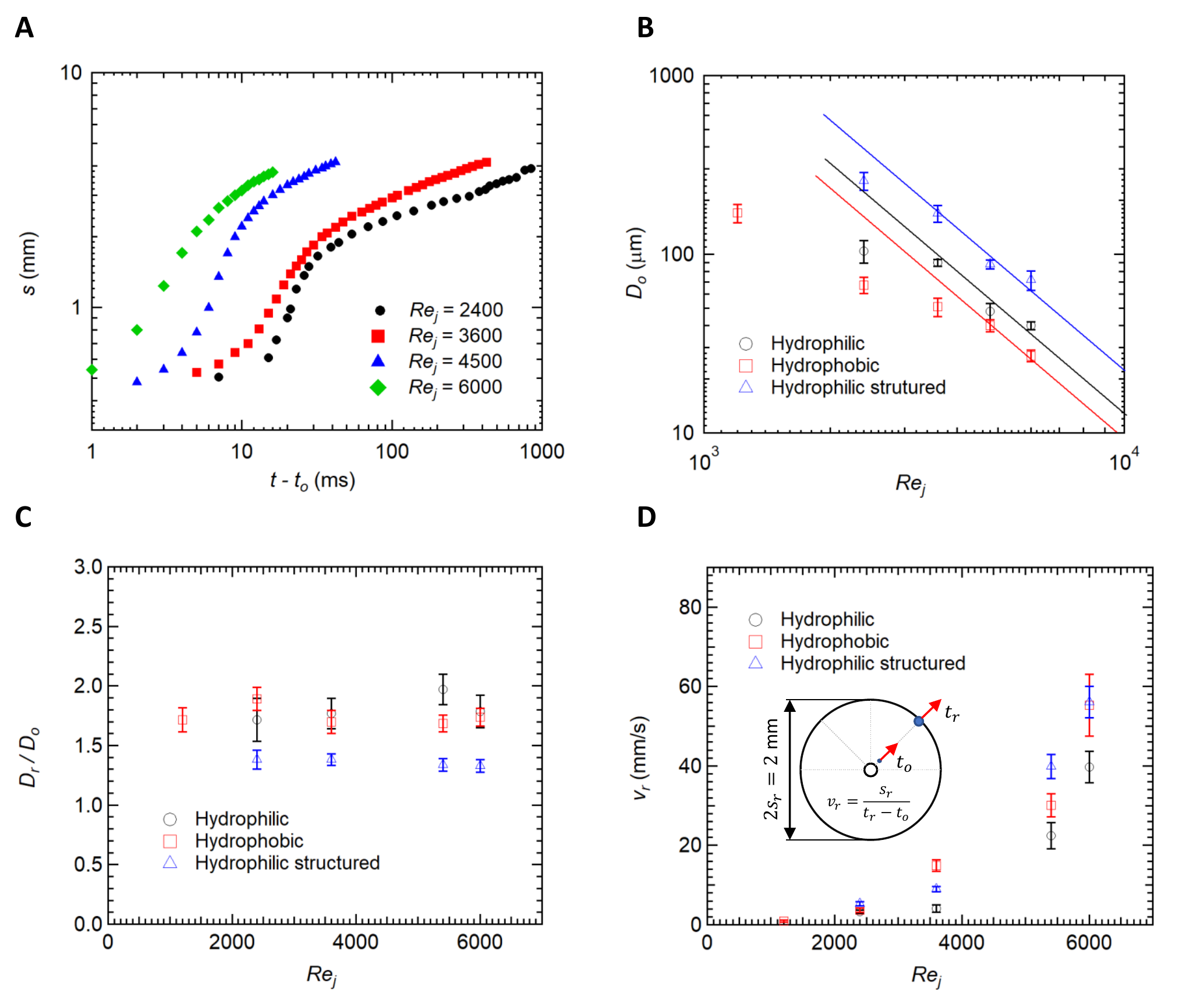}
\caption{\textbf{Dynamics of droplet-jet interaction.} \textbf{A.} Droplet location from the center of the impingement region ($s$) as a function of time ($t-t_o$), where $t_o$ is the time of onset of motion. The plot shows the effect of different jet Reynolds numbers ($Re_j$) for the hydrophobic surface ($\theta_A$ = 107$^o$ and $\theta_R$ = 103$^o$). \textbf{B.} The diameter of droplets at the onset of droplet departure ($D_o$) is depicted for three different surfaces at different jet Reynolds numbers ($Re_j$). \textbf{C.} For a comparison purpose, ratio of droplet diameter at a reference location ($s_r = 1$ mm) from the center of the impingement region to that at the onset of motion ($D/D_o$) is plotted against the jet Reynolds numbers ($Re_j$) for three different surfaces. \textbf{D.} a reference apparent speed of droplets ($v_r=s_r/(t_r - t_o)$) at a reference location ($s_r = 1$ mm) is plotted against the jet Reynolds numbers ($Re_j$) for three different surfaces.}
\label{Fig7}
\end{figure}

In Fig.7A, the droplet location ($s$) is shown as a function of elapsed time ($t-t_o$) for four different jet Reynolds numbers ($Re_j$) for the hydrophobic surface, where $t_o$ is the time of onset of motion. The different curves show similar trends depicted in Fig.6B and the existence of three different motion periods. Note that the transition from period II to period III is delayed by lowering the jet Reynolds number. Additionally, the droplet begins to decelerate at a location further from the center of the impinging jet for larger jet Reynolds numbers. Ideally, we would like to collapse all of these curves onto a single master curve.  The obvious choice is non-dimesionalize the time on the x-axis with the jet velocity divided by the jet diameter, $(t-t_0)V_j/D_j$.  Unfortunately, this simple non-dimensionalization does not collapse the data.  Nor will the data collapse if the critical diameter for droplet motion, $D_0$, is used in place of the jet diameter, $(t-t_0)V_j/D_0$.  The dynamics of drop motion are clearly quite complicated in this process.

In Fig.7B, the diameter of droplets is presented at the onset of departure over a range of jet Reynolds numbers for three different surfaces: hydrophobic, hydrophilic and microstructured hydrophilic. For the same jet Reynolds number, the hydrophobic surface, which has the largest advancing contact line and smallest contact angle hysteresis, consistently shows the smallest critical droplet diameter for the onset of droplet motion. With decreasing advancing contact angle and contact angle hysteresis, the critical droplet diameter increases for a given jet Reynolds number with the microstructured hydrophilic surface consistently showing the largest values of critical diameter needed to induce droplet motion.  Importantly, the critical droplet diameter can also be tuned by controlling the jet parameters, namely the jet velocity or Reynolds number. As can be seen in Fig.7B, increasing jet Reynolds number results in a significant decrease in critical droplet diameter.  At large values of jet Reynolds number, a distinct scaling can be observed that suggests, $D_0 \sim Re_j^{-2}$.  To investigate this further, forces acting on the droplet at force equilibrium are analyzed just before droplet departure. 

The dominating forces are surface tension force along the droplet's contact line and aerodynamic drag forces on the droplet surface. Viscous friction is assumed to be negligible because the droplet is static at this period.  The surface tension force along the contact line can be written as \cite{furmidge1962studies}

\begin{equation}\label{FD}
    F_s = \Gamma_L D_o \gamma (\cos \theta_R  - \cos \theta_A)
\end{equation}

where $\Gamma_L$ is a contact line shape factor and $\gamma$ is the interfacial surface tension. The value of the contact line shape factor depends on the droplet geometry and symmetry. In general, its value has been experimentally found to lie within $1/2$ and $\pi/2$. \cite{extrand1990retention,elsherbini2006retention} It has been observed that the state of motion of a droplet affects the surface tension force acting on a droplet, e.g. static and dynamic behavior \cite{gao2018drops}. Additionally, external vertical forces on the droplet can play a role in pinning the droplet \cite{tadmor2009measurement}. Rather than resorting to a different mathematical description as suggested by Tadmor et al. \cite{tadmor2011approaches}, Eq.\ref{FD} is used and the mentioned effects are absorbed in the order one contact line shape factor, $\Gamma_L$.

At equilibrium, the contact line force is balanced by the aerodynamics force generated by impinging jet that is trying to dislodge the droplet. The effects of velocity variation across the boundary layer can be ignored as the droplets are significantly larger than the boundary layer thickness (see supplementary material section S-5) \cite{furmidge1962studies,dussan1985ability,dussan1987ability,roisman2015dislodging}. Thus, the drag force, which is proportional to the stagnation pressure on the droplet, can be written as  
\begin{equation} \label{drag}
    F_d = \frac{1}{8} \rho_g \Gamma_A C_d u_o^2 D_o^2 
\end{equation}
where $C_d$ is the drag coefficient of the droplet, $\rho_g$ is the surrounding gas density, $\Gamma_A$ is a shape factor of the projected area of the droplet in the flow direction (see supplementary material section S-5), and $u_o$ is the effective air velocity around the droplet.  Note that $u_o$ scales with the jet mean velocity, $u_o \sim v_j$, and decays with distance from the location of jet impingement.  Balancing the forces acting on a droplet at the onset of motion, yields the following relation.
\begin{equation} \label{equilibrium}
    D_o = \frac{ \gamma (\cos \theta_R  - \cos \theta_A) \Gamma_L}{ \frac{1}{2} \rho_g C_d \Gamma_A} \frac{1}{v_j^2} \sim \frac{1}{Re_j^2}
\end{equation}
Note that the scaling of critical diameter with jet Reynolds number predicted by Eq. \ref{equilibrium} fits the experimental data in Fig.7B quite well for the high jet Reynolds number cases of all three surfaces tested. However, a deviation is observed at the lowest jet Reynolds number studied. This is likely attributable to the complexity of evaluating the geometrical shape factors and the drag coefficient which are not constant as assumed her, but depend on the air velocity and diameter of the drop \cite{milne2009drop}.

Following the onset of droplet motion, the droplets grow through a combination of continued condensation, which is slow, and coalescence with smaller stationary droplets in their path, which is fast.  At least initially, increasing droplet size is observed to result in an acceleration of the drop and an increased droplet velocity. As shown by Eq. \ref{FD}, the interfacial retention force increases with $F_s \sim D$, while the aerodynamic drag force in Eq. \ref{drag} scales increases with $F_d \sim D^2$.  Hence, as the droplets coalesce and grow beyond the critical diameter for droplet motion where these two forces are in balance, a force imbalance favoring aerodynamic drag over retention force exists and the droplets accelerate.  In order to quantify the rate of droplet diameter growth during this acceleration period, the droplet diameter normalized by the critical diameter, $D_r/D_o$, was measured at a reference location ($s_r = 1$ mm) and is plotted against the jet Reynolds number for each of the three surfaces. A schematic diagram is shown in  in Fig.7D as a reference. The reference location lies within the accelerating period (Period II) for all the cases presented. In Fig.7C, the diameter of each droplets was found to increases significantly by the time it reached a position 1mm from the center of the impinging jet.  The hydrophilic structured surface, which had the largest contact angle hysteresis shows the smallest increase in droplet diameter with an average of just $D_r/D_o = 1.35$.  This diameter increase suggests a volume increase of roughly 2.5x meaning that on average 2.5 coalescence events with similar sized drops have occured during the first 1mm of travel.  Within uncertainty, no significant difference between the hydrophilic and the hydrophobic cases could be observed.  On average, the droplets on these surface grew faster than the microstructured hydrophilic surface with an increase of in droplet diameter of roughly $D_r/D_o = 1.7$ and a corresponding volume change of 5x. It is interesting to note that even though the size of droplets is different under varying jet Reynolds numbers, the ratio of increase is constant for the same surface wettability independent of jet Reynolds number. This suggests a self-similarity property of the condensation process.

In Fig.7D, the velocity of the droplets, $v_r$ is presented as a function of jet Reynolds number.  Here, again the velocity is measured at a location within the accelerating region ($s_r = 1$ mm).  The droplet velocity can be seem to vary monotonically with the jet Reynolds number. Moreover, the droplet apparent speeds are similar for the three surfaces within the experimental uncertainty. A simplistic equation of motion (EOM) is given in supplementary material section S.6. In the acceleration period, the drag force is assumed to be much greater than the retention forces. This in turn means that the droplet motion in Period II should be independent of the surface wettability which agrees with our experimental observation in Fig 7D.

\section*{Discussion}

A novel continuous drop-wise condensation that is efficient and compact was achieved by utilizing jet impingement of water vapor on a cooled surface. The fluid dynamics of an impinged jet showed an excellent shedding capability to overcome the limitations of the state-of-the-art techniques. 

In summary, we have demonstrated the capability of our CDC design in improving the condensation process substantially compared to state-of-the-art condensers and humidification technologies. We performed condensation experiments on modified and unmodified Silicon substrates on a broad range of contact angles. The drop size being shed was controlled by tuning the jet parameters namely the jet velocity in this work. We showed that micron-sized droplets could be shed effectively even on hydrophilic surfaces. 

By comparing the condensation rate per unit volume of state-of-the-art dehumidification technologies, we showed that our design is at least six fold higher. This significant increase is attributed to the thinning of the diffusion layer which is known to impede vapor condensation. We also illustrated that by controlling the maximum droplet size being shed, improvements as high as 375\% in heat flux was possible for steam condensers in the absence of NCG. 

Finally, we discussed the droplet dynamics and growth under the jet impingement action. By comparing the different forces acting on a droplet, we were able to predict the size of droplet being shed under varying jet Reynolds number and surface wettability. Furthermore, the models presented in this work are the starting point for further optimization of the design to obtain more compact dehumidification using CDC technique.   

\subsection*{Further discussion}

The droplets generated withing the impingement area are of sizes less than the capillary length of water (or bond number $Bo << 0.1$). Therefore, gravitational force effect is negligible which in turn means the current analysis is independent of surface orientation. Gravitational force will be effective on the stationary droplets which reside in the region corresponding to the decelerating period (Period III, Fig.6B). The effect of gravitational force as well as other dominating forces, such as viscous dissipation are subjects of future research.   


It was noticed that in surfaces that are characterized by high contact angle hystereses droplets tend to stop at random locations corresponding to the decelerating period ( Period III in Fig.6B). This indicates that the droplet shedding advantage of jet impingement is lost away from the impingement region. We believe that this does not undermine the effectiveness of CDC as most of vapor condensation takes place where shedding is significant. To overcome the accumulation of condensate in that region, different engineering solutions could be applied. Placing the condensation surface vertically helps with shedding larger droplets. It is experimentally observed that stationary droplets in regions corresponding to period III grow quickly by droplet feeding coming from the impingement region. Other methods, such as surface texturing or capillary wicking could be utilized to effectively drain excessive condensate.     

Surfaces with low contact angle hystereses tend to show better uniformity of droplet shedding and stoppage location. The shedding of droplets is further improved by droplet jumping in the case of superhydrophobic surface ($\theta_A$ = 157$^o$ and $\theta_R$ = 154$^o$). As noticed in Fig.3E and video S.6, droplets in the micro-scale jump off the surface due to the release of energy upon coalescence. This phenomenon has been shown earlier to have potential in improving DWC \cite{miljkovic2012jumping} and energy harvesting \cite{Miljkovic2014}. A great advantage offered by CDC is that jumping droplets are further shed by flow generated from jet impingement. Hence, mitigate the return of micro-droplets on the surface and therefore prevent flooding of the surface.

\noindent \textbf{Supplementary Material} are presented in the attached file.

\section*{Materials and Methods}

\subsection*{Surface preparation and characterization}

Commercially available Silicon wafers (Techgophers) were used as the base condensation surfaces. In this work, we utilized five surface modifications namely (1) Hydrophilic Si surface, (2) hydrophilic micro-structured Si surface, and (3) hydrophobic silanized Si surface (4) superhydrophobic micro-structured Si surface, and (5) superhydrophobic spray-coated Si surface. Silicon wafers of similar thicknesses were used in this work to provide similar thermal resistances over the different tests. Checking the temperature distribution on the Si surfaces shows that their thermal resistance is quite negligible compared to the vapor-NCG side dominating thermal resistance. Before each experiment, the surfaces were cleaned with acetone (J.T.Baker), isopropanol (J.T.Baker), ethanol (J.T.Baker) and DI water and dried with filtered nitrogen stream. Experiments were done immediately after the cleaning process so that the effect of organic compounds found in room environment is negligible.

After Si wafers were cleaned different methods were utilized to change their wettability. The untreated Si wafer provides the base surface which happens to be slightly hydrophilic ($\theta_A$ = 85$^o$ and $\theta_R$ = 72$^o$). For altering the surface wettability, extra steps were performed other than the cleaning process. Micro-posts are fabricated on top of the silicon wafer with a diameter of 34 $\mu$m, height of 34 $\mu$m, and pitch of 50 $\mu$m. Because water condensation exists between the posts, a Wenzel state is noticed which renders the silicon wafer more hydrophilic ($\theta_A$ = 70$^o$ and $\theta_R$ = 50$^o$). To render the substrate hydrophobic ($\theta_A$ = 107$^o$ and $\theta_R$ = 103$^o$), the surface was silanized with a thin layer of polydimethylsiloxane (PDMS) using a conventional dip coating method. The film thickness is negligible compared to the thickness of the silicon wafer and therefore does not impede the heat transfer rate. Inverted micro-posts are fabricated on top of the silicon wafer with a diameter of 40 $\mu$m, height 40 $\mu$m, and pitch of 50 $\mu$m. To obtain superhydrophibicity, a combination of silanization process with the roughened surface yielded a superhydrophobic surface with a high contact angle hysteresis ($\theta_A$ = 160$^o$ and $\theta_R$ = 127$^o$). For the last surface, we utilized an aerosol spray coating method to coat the silicon wafer with WX2100 (purchased from cytonix) in which Fluorothane is the active ingredient. The resultant surface is superhydrophobic with negligible contact angle hysteresis ($\theta_A$ = 157$^o$ and $\theta_R$ = 154$^o$). The contact angle is independent of the coating thickness. Therefore, we coated the surface with about 50 $\mu$m without significant impedance of heat transfer.

The surface wettability was characterized before and after each experimental run to assess the homogeneity of their wettability as well as the consistency of contact angle measurements. Droplet shape analyzer (KR$\ddot{\rm{U}}$SS, DSA100) was used for contact angle measurements. The static advancing and receding contact angles were measured using the protocol outlined in this paper \cite{huhtamaki2018surface}. Measurements were repeated on different spots of the Si wafers to ensure homogeneity and consistency. Droplets with diameters less than the capillary length were tested to ensure negligible effects of gravity. Static advancing and receding contact angles are summarized in Table 1.

\clearpage
\subsection*{Condensation experiments}

In Fig. S1, we show a schematic of the experimental setup which consists of a bubble humidifier, a flow system, and a cooled surface. Dry filtered air was bubbled into a pool of room-temperature DI water through several spargers (Ferroday). The spargers generate micron-sized bubbles which due to their high contact area with water get humidified to above 95\%. The humid air generated exists at a room temperature ($T_\infty$ = 21$^o$C $\pm$ 1 $^o$C) and ambient humidity of (60\% $\pm$1\%). Humidity of ambient air and humidified jet was measured using a Hygrometer (VWR). The humidified air was led through a tube (Mc-MASTER-CARR) of inner diameter ($D$ = 0.047 in) to impinge normally on the cooled surface. For experimental convenience, the tube was bend to 90$^o$ while allowing enough length ($L$ = 0.84 in) before the exist section ensuring fully developed flow beyond the secondary flow region. The flow rate of humidified air was controlled by flow-adjustment valve and measured using a rotameter (OMEGA, model no. FMA-A2323). Volumetric flow rates tested range from 1 LPM to 5 LPM. The corresponding jet mean velocities range from 15 m/s to 75 m/s. The humid air jet exits the tube at a standoff distance ($H$ = 0.32 in).  

The condensation surface was the different Silicon wafers described earlier. The surfaces were placed on an Aluminum substrate with a thermally conductive paste in between. The Aluminum substrate was placed on the cold side of a Peltier plate with a thermally conductive paste. A simple peltier plate with a temperature controller unit was used to maintain a constant surface temperature ($T_s$ = 15$^o$C $\pm$ 1 $^o$C). An Infra-red (IR) camera (FLIR, A6753sc), and two flush-mounted k-type thermocouples (OMEGA, HH378) were used to observe the condensation substrate temperature as well as the condensate droplets. The substrate temperatures measured by the three sensors were in agreement within 1 $^o$C. This ruled out any possible temperature variation on the surface and ensured that the thermal resistance of vapor-gas side was dominant.

Systematic experiments were performed by first adjusting the flow to the desired jet Reynolds number $Re_j = 4 Q / \pi \nu D$, where $\nu$ is the kinematic viscosity of humid air. Then, the surface temperature was set to the desired temperature. The condensation process was allowed to reach a quasi-steady state by waiting for about 15 minutes before taking experimental measurements. To visualize the condensation process an optical microscope (Nikon, AZ100) with a high-speed camera (Photron, FASTCAM Nova) were used.

\bibliography{Manuscript}
\bibliographystyle{ScienceAdvances}

\noindent \textbf{Acknowledgements:}

A. Alshehri would like to express his sincere gratitude to King Fahd University of Petroleum and Minerals (KFUPM), Dhahran, Saudi Arabia.\\
\noindent \textbf{Funding:} This research was supported by the National Science Foundation under Grant No. CBET-2032533.\\
\noindent \textbf{Competing Interests} The authors declare that they have no competing financial interests.\\
\noindent \textbf{Data and materials availability:} Additional data and materials are available online.


\beginsupplement

\begin{figure}[p!]
\centering
\includegraphics[width=0.8\textwidth]{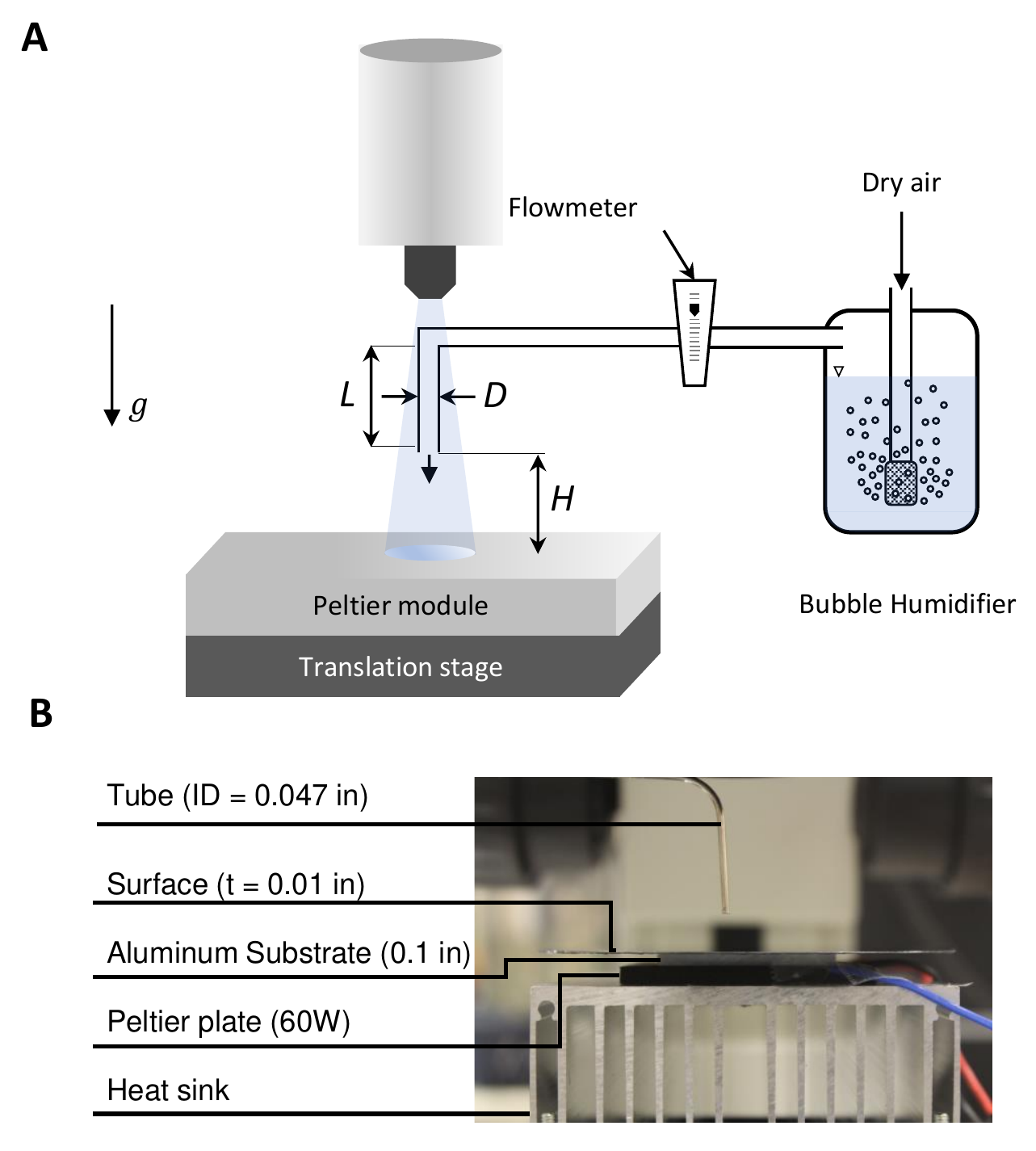}
\caption{ \textbf{ Continuous Drop-wise Condensation experimental setup.} \textbf{A.} A schematic of CDC illustrating the general setup for condensation experiments. Dry air is bubbled into a room-temperature pool of DI water through several spargers (one is shown for illustration). The different parameters are discussed in the method section. \textbf{B.} A side view of the condensation surface assembly.}
\label{FigS1}
\end{figure}

\clearpage
\section{Heat Transfer of water Vapor Condensation with Humid Air Jet Impingement}

The previous section utilized a theoretical approach to understand and quantify the heat flux to a condensation surface in the case of pure vapor condensation, i.e. no air involved. Hence, this section is dedicated to model the problem of air existence. It has been shown that the existence of minute amounts of non-condensable gases (NCG), such as air reduces the condensation rate, and thus the heat flux, tremendously. This is due to the fact that at steady state conditions, a diffusion layer builds on top of the condensate formed on the surface. The condensation of vapor becomes mainly dominated by the mass diffusion through this layer. The thermal resistance of this layer dominates the heat transfer to the surface, therefore, Eq. \ref{q_d} is not valid to describe the system. The heat transfer from the surrounding through the droplet to the condensation surface can be estimated simply as.
\begin{equation}
        q'' (r) = \Dot{m}'' h_{fg} + h (T_\infty - T_s)
\end{equation}
where the mass flux to the surface can be written as
\begin{equation}
    \Dot{m}''= \frac{\phi Sh \rho D_{ij}}{D}\bigg(\frac{\omega_{v,\infty} - \omega_{v,o}}{1-\omega_{v,o}}\bigg)
\end{equation}
where the suction effect factor is written as 
\begin{equation} \label{Suction}
\phi = \frac{1-\omega_{v,o}}{\omega_{v,o}-\omega_{v,\infty}} ln \bigg(\frac{1-\omega_{v,\infty}}{1-\omega_{v,o}} \bigg)
\end{equation}
The heat transfer coefficient can be written as
\begin{equation}
    h = \frac{Nu k}{D}
\end{equation}
Sherwood and Nusselt numbers for jet impingement can be obtained by utilizing heat/mass transfer analogy along with the correlation usually used for single round jets \cite{martin1977,Incropera2011}.
\begin{equation}\label{jet}
    \frac{Sh}{Sc^{0.42}} = \frac{Nu}{Pr^{0.42}} =  G(A_r, \frac{H}{D}) [2 Re^{1/2} (1 + 0.005 Re^{0.55})^{1/2}]
\end{equation}
where
\begin{equation}
    G =  \frac{1-1.1 D/r}{1 + 0.1 (H/D - 6) D/r} \frac{D}{r}
\end{equation}

In Fig.4B, we show the resulting heat and mass fluxes for the case of a saturated humid air jet impinging on a cold surface at varying jet parameters. The inputted parameters in the correlation are; saturated environment at $T_\infty$ = 22$^o$C and $T_s$=15$^o$C at $P$=1 atm. The nozzle diameter is 1 mm located at varying distance. The impingement area is is confined withing 2.5 mm.  

\clearpage
\section{Calculation of compactness factor of different state-of-the-art condensers:}
For a condenser existing in humid air environment, one can write the heat transfer to the surface as follows
\begin{equation}
    q = h A (T_\infty - T_s)
\end{equation}
where $h$ is the heat transfer coefficient, $A$ is the exposed surface area, $T_\infty$ and $T_s$  are the ambient and surface temperatures, respectively. Similarly, the mass transfer (condensation rate) can be written as follows
\begin{equation} \label{mass_transfer}
    \Dot{m} = h_m A (\omega_\infty - \omega_s)
\end{equation}
where $h_m$ is the mass transfer coefficient, $\omega_\infty$ and $\omega_s$ are the ambient and surface vapor mass fraction, respectively. The relation between $h$ and $h_m$ can be inferred from heat/mass transfer analogy as
\begin{equation}
    \frac{h}{h_m} = \frac{Nu}{Sh} \frac{k}{\rho D} = \frac{C Re^n Pr^m k}{C Re^n Sc^m \rho D} \approx \bigg(\frac{Pr}{Sc}\bigg)^{1-m} C_{pa} = Le^{1-m} C_{pa}
\end{equation}
The value of the exponent ($m$) depends on the correlation being used. For instance, for a laminar boundary layer over a flat plate, a value of (1/3) is used\cite{sievers2013design}. In our case, the correlation given by Eq.\ref{jet} is characterized by an exponent of (0.42). 

For comparison of the current method of condensation with other condensers/dehumidifiers, we think CDC provides an extremely compact condenser. Therefore, we compare the different state-of-the-art condensers by a compactness factor which is given by the following relation
\begin{equation}
    C_h = \frac{h A}{V}
\end{equation}
\begin{equation}
    C_m = \frac{h_m A}{V}
\end{equation}
where $C_h$ and $C_m$ are the compactness factor of heat and mass transfer exchangers, respectively. The higher the value indicates a higher transfer rate per unit driving potential (temperature or vapor mass fraction) per unit volume. In Fig.4A, we show a comparison of the compactness factor for different dehumidifiers along with the current method. We reconstruct the results published by Sadeghpour et al. \cite{sadeghpour2019water}.   

\clearpage

\section{Heat Transfer of Pure Vapor with Jet Impingement}

In this section, an estimate of the heat transfer of DWC of pure vapor is presented. We utilize the theoretical model developed originally by Rose and coworkers \cite{rose2015}. The process of DWC can be summarized by the following sequence of events; initial nucleation of vapor clusters, droplet growth by condensation on the interface, droplet growth by coalescence, droplet fall off, and finally re-nucleation of droplets. 

The first process of DWC is heterogeneous condensation over active nucleation sites on the condensation substrate. The active nucleation site density ($N_s$) depends on surface topography, Fluid's thermophysical properties, and subcooling degree \cite{rose1976further,sikarwar2013mathematical}. The value of active nucleation site density is in the range of 10$^9$-10$^{15}$ m$^{-2}$ \cite{liu2015dropwise1}. For the purpose of obtaining an approximate comparison between jet- and gravity-assisted shedding mechanisms, we choose a value for $N_s = 10^{12}$ m$^{-2}$. The smallest stable droplet formed by condensation in the nucleation site is given as \cite{graham1973drop,liu2015dropwise2}
\begin{equation}
    r_{min} = \frac{2 T_{sat} \gamma}{\rho_l h_{fg} (T_{sat} - T_s) }
\label{r_min}
\end{equation}

where $T_{sat}$, $\gamma$, $\rho_l$, $h_{fg}$, $T_s$ as saturation temperature, surface tension, liquid density, latent heat of vaporization, and surface tension, respectively. It has been shown experimentally that droplets in the range $[r_{min},r_e]$ grow by direct condensation only, where the effective radius results from geometrical argument as $r_e = 1/\sqrt{4 N_s}$. Droplets with radii higher than the effective radius grow by direct condensation on their surfaces as well as by coalescence with neighboring droplets. DWC on vertical surface, i.e. typical configuration, are characterized by the existence of a maximum droplet radius of which droplets start sliding on the surface and consequently sweeping smaller droplets in their path. A balance between the weight of the droplet and the surface retention force results in the following relation of maximum radius for gravity-assisted shedding \cite{dimitrakopoulos1999gravitational}  

\begin{equation}
    r_{max,g} = \sqrt{\frac{6 \gamma \sin{\theta} (\cos{\theta_r} - \cos{\theta_a})}{\pi \rho_l g  (2 - 3 \cos{\theta} + \cos^3{\theta})}}
\end{equation}

where $\theta$, $\theta_r$, $\theta_a$ are the static, receding and advancing contact angles respectively and $g$ is the gravitational acceleration. To estimate the overall heat transfer rate to the condensation surface, several researchers have utilized the theory developed by Rose and co-workers \cite{rose2015}. The model is centered around correlating the heat transfer across a single droplet to the overall heat transfer across the entire droplets on a surface. The following formula is usually considered for the overall heat transfer rate
\begin{equation}
    q'' = \int_{r_{min}}^{r_e} q_d(r,\theta) n(r,\theta) dr + \int_{r_e}^{r_{max}} q_d(r,\theta) N(r,\theta) dr 
\end{equation}

where $q_d(r,\theta)$, $n(r,\theta)$ and $N(r,\theta)$ are the heat transfer through a single droplet of radius ($r$), the \textit{small} droplet size distribution in the range $[r_{min},r_e]$, and \textit{large} droplet size distribution in the range $[r_e,r_{max}]$, respectively. The detailed derivation of the individual parameters has been discussed in several papers beginning with the work of Rose \cite{rose2015}. 

The heat transfer across a single droplet of radius ($r$) can be represented as a combination of Laplace pressure effect due to curvature, liquid-vapor interfacial thermal resistance (Knudsen layer), conduction through the droplet body, and conduction through the condensation surface. heat transfer across a single droplet can be written as \cite{kim2011dropwise}
\begin{equation} \label{q_d}
    q_d (r,\theta) = {\pi r^2 (T_{sat} - T_s - \frac{2 T_{sat} \gamma}{\rho_l h_{fg} r })}(\frac{1}{2 h_i (1 - \cos{\theta})} + \frac{r \theta}{4 k_l \sin{\theta}} + \frac{\delta_s}{k_s \sin^2{\theta}} )^{-1}
\end{equation}
where $k_l$ and $k_s$ are the thermal conductivity of the liquid and the condensation surface, respectively and $\delta_s$ is the thickness of the condensation surface. The liquid-vapor interfacial heat transfer coefficient is given as \cite{wen1976heat}
\begin{equation}
    h_i = \frac{2 \sigma_c}{2 - \sigma_c} \sqrt{\frac{M}{2 \pi R T_s}} \frac{h_{fg}^2 \rho_v}{T_s}
\end{equation}
where $sigma_c$, $M$, $R$ and $\rho_v$ are the condensation coefficient, molecular weight of water, gas constant, and water vapor density, respectively. The size distribution of large droplets was derived experimentally and mathematically as \cite{leFevre1966theory,rose1973dropwise}
\begin{equation}
    N(r,\theta) = \frac{1}{3 \pi r_{max} r^2} \bigg(\frac{r}{r_{max}}\bigg)^{-2/3}
\end{equation}
Lastly, the population balance theory was used to derive the small droplet size distribution \cite{abu1998modeling}. The form is given as 
\begin{equation}
    n(r,\theta) = \frac{1}{3 \pi r_{max} r_e^3} \bigg(\frac{r_e}{r_{max}}\bigg)^{-2/3} \frac{r (r_e - r_{min})}{r - r_{min}} \frac{A_2 r + A_3}{ A_2 r_e + A_3} \exp{(B_1 + B_2)}
\end{equation}
where the constants are given as
\begin{subequations}
\begin{align}
        A_1 &= \frac{(T_{sat} - T_s)}{2 \rho_l h_{fg}},         \\
        A_2 &= \frac{\theta (1 - \cos{\theta}) }{4 k_l \sin{\theta}},   \\
        A_3 &= \frac{1}{2 h_i} + \frac{\delta_s (1- \cos{\theta})}{k_s \sin^2{\theta}}  \\
        A_4 &= \frac{A_2}{\tau A_1} \bigg[ \frac{r_e^2 - r^2}{2} + r_{min} (r_e - r) + r_{min}^2 \ln{\bigg(\frac{r_e - r_{min}}{r - r_{min}}\bigg)} \bigg]  \\
        A_5 &= \frac{A_3}{\tau A_1} \bigg[r_e -r + r_{min} \ln{\bigg( \frac{r_e - r_{min}}{r - r_{min}} \bigg)} \bigg] \\
        \tau &= \frac{3 r_e^2 (A_2 r_e + A_3)^2}{A_1 [8 A_3 r_e - 14 A_2 r_e r_{min} + 11 A_2 r_e^2 - 11 A_3 r_{min}]}
\end{align}
\label{n_small}
\end{subequations}
Equations \ref{r_min} through \ref{n_small} along with known thermophysical properties provide a complete set of equations to obtain the overall heat transfer rate to a surface. For instance, Fig.S3 shows the heat flux variation on a surface with different wettability, i.e. contact angles, under gravity-assisted droplet shedding. In the figure, we also show the maximum radius of droplets being shed by the assistance of body weight. For a surface with static contact angle of 90$^o$, we notice that a heat flux of 30 kW/m$^2$ could be transferred to the surface with a maximum droplet radius of 390 microns. With higher contact angle surfaces, droplets of lower radius could be shed. However, due to the increased conduction resistance, heat flux drops beyond a contact angle of 145$^o$.

In Fig.S4, we show for a constant static contact angle, the effect of varying the maximum droplet radius. The cross symbols in the figure represent the case of gravity-assisted droplet shedding as their respective static contact angles. For instance, the heat flux to a surface could be enhanced by 150\% if a coating with static contact angle of 160$^o$ is used with a maximum droplet radius of around 20 micron compared with the hydrophilic surface and gravity-assisted case. 

We believe that with CDC, the mechanism of droplet growth would be similar to that of regular drop-wise condensation with the maximum droplet radius determined by the jet impingement shedding action. Therefore, in Fig.4D we normalize the heat flux and maximum droplet radius with with their respective values due to gravity-assisted drop-wise condensation.

\begin{figure}
\centering
\includegraphics[width=0.7\textwidth]{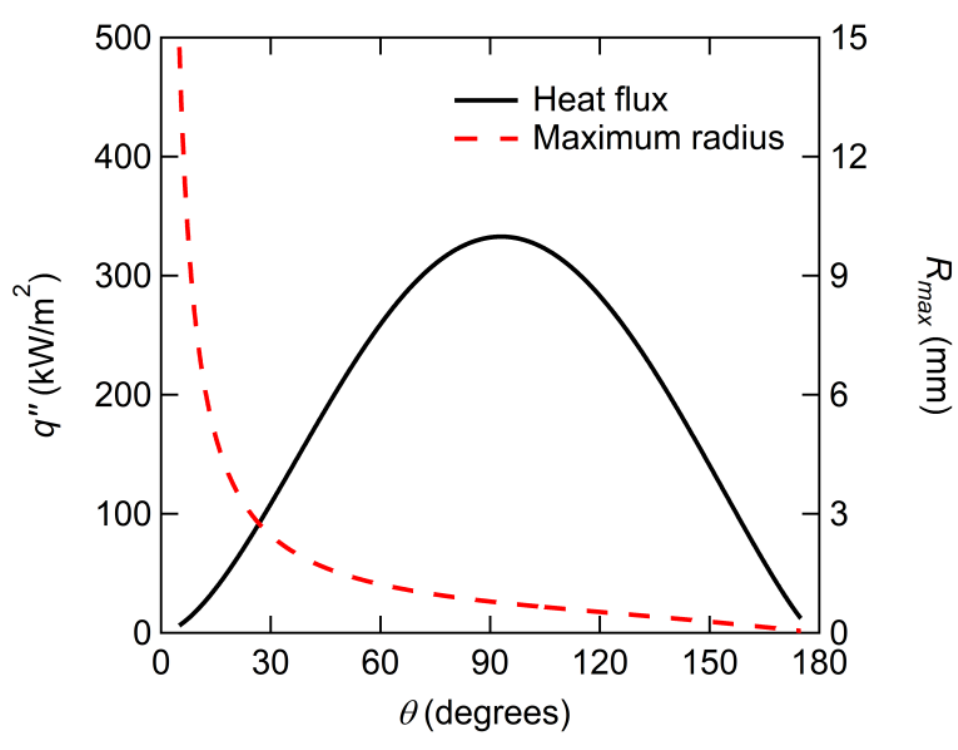}
\caption{ Heat flux to a vertical surface exposed to \underline{pure vapor} analytically evaluated at different static contact angles with gravity-assisted shedding. The parameters inputted into the model are; $T_{sat}$=22$^o$C, $T_s$=15$^o$C, $\delta_s$ = 254$\mu$m, $k_s$=100W/m$^2$K, $N_s$=10$^{12}$ sites/m$^2$, $\sigma_c$=1, and $\theta_A - \theta_R$=5$^o$. }
\label{FigS3}
\end{figure}

\begin{figure}
\centering
\includegraphics[width=0.7\textwidth]{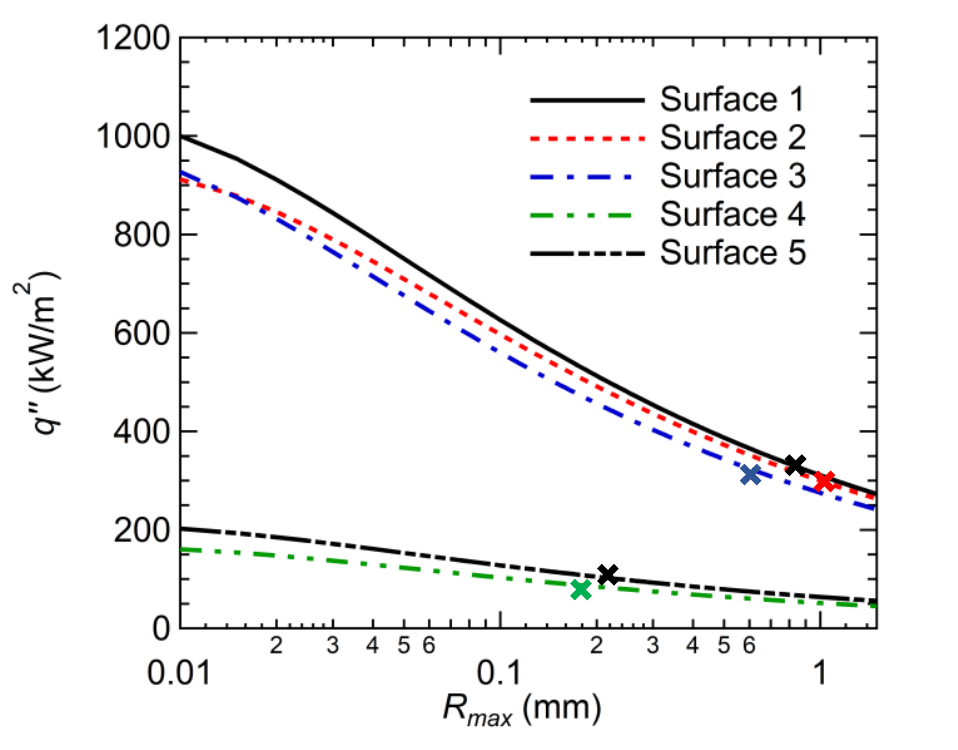}
\caption{ Heat flux to a vertical surface exposed to \underline{pure vapor} analytically evaluated at maximum droplet Radii for the different tested surfaces. The cross symbol represents the value obtained with gravity-assisted shedding. The parameters inputted into the model are similar to that in Fig.S5 for meaningful comparison.}
\label{FigS4}
\end{figure}

\clearpage

\section{ Discussion of drag force quantification }

As observed in (Fig.1 and Fig.2 of main manuscript), the initial droplet shedding occurs withing the stagnation region, i.e. almost within a tube radius from the stagnation point. It is well established that for the case of laminar non-mixing gas jet impingement, the similarity solution of Navier-Stokes equations gives a constant momentum boundary layer thickness in the impingement region. The equation of the boundary layer thickness is usually written as $\delta_o \sim \sqrt{\nu D / u_j} $, where $\nu$ is the kinematic viscosity of the jet, $D$ is the diameter of the tube exit, and $u_j$ is the speed of the jet at the standoff distance \cite{schlichting1968,lienhard1995liquid}. In this work, the jet Reynolds number was in the range of 1000-6000, therefore, the momentum boundary layer thickness should be below the range of 15 - 35 $\mu m$, respectively. Even though these estimations are for laminar jets ($Re<1000$), turbulent mixing, like in our experiments, results in a lower boundary layer thickness. Our observations of droplets size show that the boundary layer thickness is smaller than the smallest droplet being shed. Therefore, we expect that the flow field withing the boundary layer is unimportant. 

In general, the drag force on the droplet due to the jet flow can be given by equation 7. The projected area shape factor $\Gamma_A$ can be obtained geometrically as
\begin{equation}
    \Gamma_A = \theta - \frac{1}{2} \sin 2 \theta
\end{equation}

The drag coefficient based on the jet speed and droplet diameter is in the range of 1-0.6 for Reynolds numbers of 100-500, respectively. These values were obtained from spherical relations for the lack of better quantification in the literature, however it is a common practice. The effective velocity term is taken to equal the jet mean velocity value multiplied by a proportionality constant, i.e. $u_o = a v_j = Q/A_{tube}$.

\clearpage

\section{ Droplet equation of motion: simplistic approach }

In this section, we present a simplistic equation of motion that represents a one-dimensional force balance on a single droplet. In Fig.S5A, we show the simplified physical model of a droplet with diameter ($D$) and contact angle ($\theta$) located at a distance ($x_o$) from the center of impingement region. The jet issues from a tube that is located at a standoff distance ($H$) with a mean velocity ($v_j$). The only forces responsible for droplet movement are drag force ($F_d$), surface tension force ($F_s$), and viscous friction force ($F_v$) (Fig.S5B). Newtons seconds law is applied to the droplet as follows.
\begin{equation}\label{EOM}
    \rho_l \Gamma_v \frac{d(D^3 v)}{dt} = \frac{1}{8} C_d \rho_g \Gamma_A (u_o - v)^2 D^2 - \Gamma_L D \gamma (\cos \theta_R  - \cos \theta_A) - \eta \Gamma_{base} R^2 \frac{\partial v}{\partial y}\bigg|_{base}
\end{equation}
where $\rho_l$ is the density of condensate liquid, $v$ is the local velocity of the droplet, $\Gamma_v$ is the volumetric shape factor, $\eta$ is the dynamic viscosity, $\Gamma_{base}$ is the shape factor of the base area of the drop. At the onset of droplet motion, only the drag and surface tension forces are present. Equation 8 in the main manuscript is the resultant of the force balance. In the accelerating droplet region (Period II), the drag force becomes significantly higher than the retention forces. Hence, Eq.\ref{EOM} is reduced to the following.
\begin{equation}
    \rho_l \Gamma_v \frac{d(D^3 v)}{dt} = \frac{1}{8} C_d \rho_g \Gamma_A (u_o)^2 D^2
\end{equation}
This equation suggests a negligible effect of surface wettability on droplet motion in regions corresponding to Period II. This is clear from the droplet apparent velocity shown in Fig.6D. In the decelerating period (Period III), Eq.\ref{EOM} should be fully used. However, because of the complexity of determining the coefficients in the equation, presenting an outline of the equation is sufficient.

\begin{figure}[h]
\centering
\includegraphics[width=0.7\textwidth]{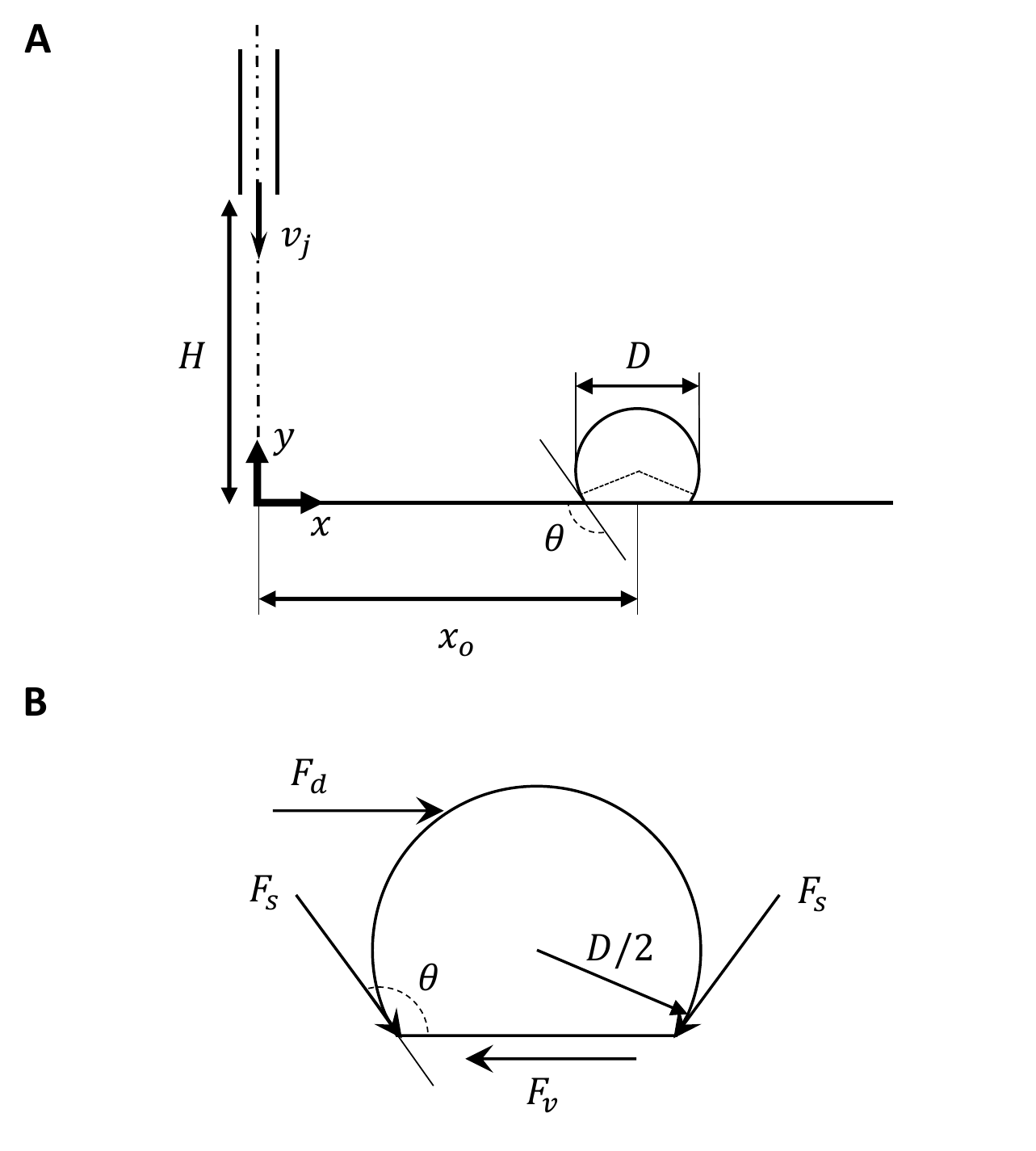}
\caption{ Physical model for writing the equation of motion of a single droplet in contact with flow of an axisymmetric jet. \textbf{A.} one-dimensional schematics of the pertaining parameters. \textbf{B.} Forces acting on a single droplet under a generalized case of a moving droplet.}
\label{FigS5}
\end{figure}

\clearpage

\section{Videos}

\paragraph{Video S1} 10-second Condensation process on surface 1 (hydrophilic Silicon surface) at different jet Reynolds numbers. The experimental conditions are as indicated in the method section while surface characteristics are as in Table S1.

\paragraph{Video S2} 10-second Condensation process on surface 2 (hydrophilic silicon surface with high contact angle hysteresis) at different jet Reynolds numbers. The experimental conditions are as indicated in the method section while surface characteristics are as in Table S1.

\paragraph{Video S3} 10-second Condensation process on surface 3 (hydrophobic Silicon surface) at different jet Reynolds numbers. The experimental conditions are as indicated in the method section while surface characteristics are as in Table S1.

\paragraph{Video S4} 10-second Condensation process on surface 4 (superhydrophobic silicon surface with high contact angle hysteresis) at different jet Reynolds numbers. The experimental conditions are as indicated in the method section while surface characteristics are as in Table S1.

\paragraph{Video S5} 10-second Condensation process on surface 5 (superhydrophobic silicon surface with negligible contact angle hysteresis) at different jet Reynolds numbers. The experimental conditions are as indicated in the method section while surface characteristics are as in Table S1.

\paragraph{Video S6} Illustration of droplet jumping on surface 5 (superhydrophobic silicon surface with negligible contact angle hysteresis).

\end{document}